%% file: GluonPaper.tex
\documentclass[12pt]{article}

\pdfoutput=1
\usepackage{color}
\usepackage{epsfig, palatino}
\usepackage{pstricks,pst-node,pst-tree}
\usepackage{epic}
\usepackage{mathrsfs}
\usepackage{ae} 
\usepackage[T1]{fontenc}
\usepackage[ansinew]{inputenc}
\usepackage{amsmath}
\usepackage{amssymb}
\usepackage{graphicx}
\usepackage{ulem}
\usepackage{color}
\definecolor{darkblue}{cmyk}{0.9,0.9,0,0}
\usepackage[colorlinks=true,linkcolor=darkblue,citecolor=darkblue,urlcolor=darkblue]{hyperref}
\usepackage{cite}
\usepackage{hyperref}
\usepackage{wasysym}
\usepackage{varioref}
\usepackage{makeidx}
\usepackage[english]{babel}
\usepackage{simplewick}
\usepackage{array}

\newcommand{\comment}[1]{}

\newcommand{\beq}{\begin{equation}}
\newcommand{\eeq}{\end{equation}}
\newcommand{\beqq}{\begin{equation*}}
\newcommand{\eeqq}{\end{equation*}}
\newcommand\beqa{\begin{eqnarray}}
\newcommand\eeqa{\end{eqnarray}}
\newcommand\beqaa{\begin{eqnarray*}}
\newcommand\eeqaa{\end{eqnarray*}}
\newcommand\bea{\begin{array}}
\newcommand\eea{\end{array}}

\newcommand{\nn}{\nonumber}

\newcommand{\neqa}{\nonumber\end{eqnarray}} 
\newcommand{\la}[1]{\label{#1}}

\renewcommand{\d}{\partial}

\newcommand{\<}{{\langle}}
\renewcommand{\>}{{\rangle}}

\newcommand{\re}{\relax{\rm I\kern-.18em R}}

\renewcommand{\sp}{p\hspace{-.40em}/}

\newcommand{\ft}[2]{{\textstyle\frac{#1}{#2}}}

\newcommand{\Blue}[1]{{\color{blue}#1\color{black}}}
\newcommand{\Red}[1]{{\color{red}#1\color{black}}}

\def\XXint#1#2#3{{\setbox0=\hbox{$#1{#2#3}{\int}$}
\vcenter{\hbox{$#2#3$}}\kern-.5\wd0}}

\def\su2{{SU(2)}}

\def\a{{\alpha}}

\def\[{\left[}
\def\]{\right]}

\def\a{\alpha}
\def\b{\Bethe}

\def\({\left(}
\def\){\right)}
\def\[{\left[}
\def\]{\right]}

\def\<{\langle}
\def\>{\rangle}

\def\i2{\frac{i}{2}}

\def\spi{\relax{\rm \pi\kern-0.5em /}}
\def\sA{\relax{\rm A\kern-0.5em /}}
\def\sp{\relax{\rm p\kern-0.5em /}}
\def\sd{\relax{\rm \d\kern-0.5em /}}
\def\sk{\relax{\rm k\kern-0.5em /}}
\def\sn{\relax{\rm n\kern-0.5em /}}
\def\sl{\relax{\rm l\kern-0.5em /}}
\def\sP{\relax{\rm P\kern-0.7em /}}
\def\sBethe{\relax{\rm \Bethe\kern-0.5em /}}

\def\cW{{\cal W}}

\def\2F1{\,_2{\rm F}_1}

\makeindex

\begin{document}

\thispagestyle{empty}

\renewcommand{\thefootnote}{\fnsymbol{footnote}}
\setcounter{page}{1}
\setcounter{footnote}{0}
\setcounter{figure}{0}
\begin{center}
$$$$
{\Large\textbf{\mathversion{bold}
Space-time S-matrix and Flux-tube S-matrix IV.\\ 
Gluons and Fusion
}\par}

\vspace{1.0cm}

\textrm{Benjamin Basso$^{{\displaystyle\hexagon}}$, Amit Sever$^{{\displaystyle\Box}}$ and Pedro Vieira$^{\displaystyle\pentagon}$}
\\ \vspace{1.2cm}
\footnotesize{\textit{
$^{\displaystyle\pentagon}$Perimeter Institute for Theoretical Physics,
Waterloo, Ontario N2L 2Y5, Canada\\
$^{\displaystyle\hexagon}$Laboratoire de Physique Th\'eorique, \'Ecole Normale Sup\'erieure, Paris 75005, France\\
$^{\displaystyle\Box}$School of Natural Sciences, Institute for Advanced Study, Princeton, NJ 08540, USA
}  

\vspace{4mm}
}

\par\vspace{1.5cm}

\textbf{Abstract}\vspace{2mm}
\end{center}
We analyze the pentagon transitions involving arbitrarily many flux-tube gluonic excitations and bound states thereof in planar $\mathcal{N}=4$ Super-Yang-Mills theory. We derive all-loop expressions for all these transitions by factorization and fusion of the elementary transitions for the lightest gluonic excitations conjectured in a previous paper. We apply the proposals so obtained to the computation of MHV and NMHV scattering amplitudes at any loop order and find perfect agreement with available perturbative data up to four loops.

\noindent

\setcounter{page}{1}
\renewcommand{\thefootnote}{\arabic{footnote}}
\setcounter{footnote}{0}

 \def\nref#1{{(\ref{#1})}}

\newpage

\tableofcontents

\parskip 5pt plus 1pt   \jot = 1.5ex

\section{Introduction}

In its original incarnation, String Theory was proposed as the description of the colour flux tube which holds quarks together. It was soon after dethroned by Quantum Chromodynamics as the description of the strong interaction. The current wisdom is, however, that the two descriptions actually coexist or, in more modern terms, are \textit{dual} to each other. Unfortunately we do not have a crisp exact description of the QCD string at our disposal yet.
Were it known, we would be learning about the interactions of quarks and gluons in four dimensions from the (supposedly simpler) study of the one dimensional flux tube dynamics. 

One of the most fascinating aspects of planar $\mathcal{N}=4$ SYM theory, a supersymmetric distant cousin of QCD, is the nearly absolute control over its associated colour flux tube, i.e.~of its dual string.
Not only is the dual string theory known~\cite{Maldacena:1997re} but it so happens that the flux tube dynamics is as simple as it could be: the flux tube excitations interact with each other in a factorized way. In other words, the flux tube of this gauge theory is \textit{integrable}~\cite{review} and extremely well understood \cite{AldayMaldacena,BenDispPaper,OPEpaper}.

The formalism which tames the flux tube and puts it to use in the study of gluon scattering amplitudes goes by the name of the \textit{pentagon approach}. It was proposed in \cite{short} as a refinement of the so called OPE program~\cite{OPEpaper}. In this approach, scattering amplitudes in planar $\mathcal{N}=4$ SYM theory are given by an \textit{OPE sum} over all multi-particle excitations of the flux tube. It is certainly a formidable task to spell out this sum in full detail. After all, there is a plethora of flux tube excitations, see figure \ref{truncation}, and thence a multitude of multi-particle states one should sum over.

\begin{figure}[t]
\centering
\def\svgwidth{16cm}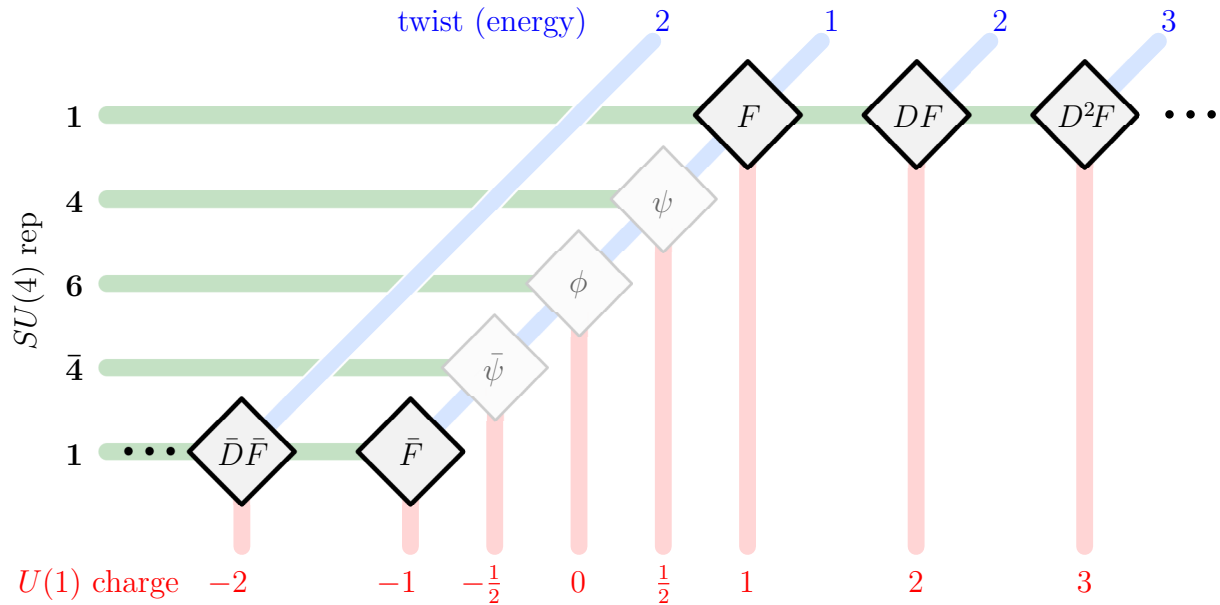
\caption{Picture of flux tube excitations and their quantum numbers. Lying on the diagonal are the twist-one excitations which can be scalar $\phi$, fermionic $\psi, \bar{\psi}$, or gluonic $F, \bar{F}$. The latter excitations can form bound states depicted on the bottom and top rows. There is precisely one bound state at any given twist $2, 3, \ldots$ and $U(1)$ charge $\pm 2, \pm 3, \ldots\,$, denoted by $DF\sim F^2, D^2F\sim F^3, \ldots$ or their complex conjugates. In this paper we consider the OPE contribution from states made out of any number of gluons and bound states, that is built out of the excitations presented in the boldfaced squares only.
} \la{truncation}
\end{figure}

There is, however, a natural hierarchy amongst all those states. The more particles we have, the heavier the state is (i.e.~the biggest is its flux tube energy) and thus the more suppressed is its contribution to the OPE sum. It is thus very natural to begin with the states with the smallest excitation numbers. This is precisely what we did in \cite{data} and \cite{2pt} where we analyzed in detail the contributions of the states with one- and two-particle respectively. 

In this paper, we initiate a more systematic study of the multi-particle states by considering the OPE contributions involving \textit{arbitrarily} many flux tube gluonic excitations, see figure \ref{truncation}.
In other words, we will drop the scalar and fermionic excitations and study all that remains.
This is clearly a very brutal truncation of the full OPE series. Still, it defines an interesting subsector for both physical and technical reasons. 

Firstly, the gluonic excitations are associated to the transverse fluctuations of the flux tube which are present in any gauge theory.
In this sense, they are the most universal amongst all flux tube excitations and the gluonic subsector stands as the most representative of all.

Furthermore, at strong coupling, scattering amplitudes are given by a minimal surface computation in $AdS_5$~\cite{AM}. From the OPE viewpoint, this result comes mostly from resumming the contributions of any number of gluon excitations~\cite{short,toappear}.\footnote{This is a slight oversimplification as explained in~\cite{2pt,O6paper} in more detail. }  This is in line with the usual AdS/CFT cartoonish intuition which associates the physics of the sphere with that of the gauge theory scalars and the physics of AdS with the gluonic dynamics of the gauge theory.
Understanding the gluons and their bound states is therefore a necessary step toward explaining how the full minimal area prescription emerges out of the OPE.  

Yet another motivation comes from the interplay with the perturbative analysis at weak coupling. In this context, the OPE is providing valuable boundary data for the scattering amplitude bootstrap program laid out in \cite{Dixon:2013eka,Dixon:2014voa}. The gluonic sector includes in particular two subsectors that have proven to be extremely useful. They correspond to truncating the OPE series to the contributions of multi-gluon states all of the same helicity (be it positive or negative) only. Intriguingly, requiring the perturbative result to match these maximal helicity contributions for one- and two-gluon states has been enough thus far to bootstrap the hexagon Wilson loop to four loops within the hexagon function program \cite{Dixon:2013eka,Dixon:2014voa}. We can not rule out the optimistic possibility that these subsectors alone -- with an arbitrary number of gluons -- might suffice to bootstrap the full hexagon at all loops. 

The last reason is more technical. The scalars and fermions, which we are disregarding here, transform non-trivially under the $SU(4)$ R-symmetry, see figure \ref{truncation}. 
As a consequence, their pentagon transitions are $SU(4)$ tensors, with as many indices as excitations involved. The gluonic transitions are free of such a complication and hence much easier to study. They serve as a laboratory for understanding the abelian components (a.k.a.~dynamical parts~\cite{data,O6paper}) of the multi-particle transitions in general.

The strategy adopted in this paper is the following. First, we will bootstrap the transitions for multi-particle states of the lightest gluonic excitations (in Section~\ref{section2}). This will allow us to make contact with the conjectures put forward in~\cite{short}. Next, we shall fuse these elementary objects  together and obtain the general transitions involving bound states as well (in Section~\ref{section3}). Finally, we shall explain how to convert these predictions into finite coupling results for scattering amplitudes. We shall focus on the MHV and NMHV 6- and 7-points amplitudes and compare our findings with the available perturbative data (in Section~\ref{section4}).

\section{Multi-particle Transitions}\label{section2}

The lightest gluonic excitations are the twist-one gluons $F$ and $\bar F$ in figure~\ref{truncation}. (In terms of the components of the Faraday tensor, 
$F=F_{-z}$ and $\bar{F}=F_{-\bar{z}}$ while in bi-spinor notation $F=\mathcal{F}_{11}$ and $\bar F=\mathcal{F}_{\dot 1\dot 1}$, see \cite{data}.) 
We also have heavier gluonic excitations which can be thought of as bound states of the lightest ones. 
In this paper we shall employ the unifying notation 
\beq 
F_{a}(u) \la{fnotation}
\eeq 
to indicate a gluonic excitation carrying rapidity $u$ (or, equivalently, momentum $p_{a}(u)$ along the flux tube direction). The index $a$ will allow us to keep record of the $U(1)$ charges of the gluon, with $a=1$ for the positive helicity twist-one gluon $F$ and $a=-1$ for its negative helicity counterpart $\bar F$. In this notation a bound state of $n$ gluons of positive/negative helicity is denoted as $F_n(u)/F_{-n}(u)$ respectively. 

Throughout we shall also use $\mu_a(u)$ to indicate the square measure of the excitation~(\ref{fnotation}). Equivalently, our states are normalized as 
\beq
\langle F_{b}(v) |   F_{a}(u)  \rangle= \frac{2\pi}{\mu_a(u)}\,\delta_{ab}\, \delta(u-v)\, ,
\eeq
and similarly for multi particles, see \cite{data} for further details. The main result of this paper is the bootstrap of the gluon pentagon transition\footnote{We shall drop the separator `$|$' in the lower indices of~$P$ if either the initial or final state is the vacuum.} 
\beq
P_{a_1,\dots, a_N |b_1,\dots, b_M} (u_1,\dots,u_N|v_1,\dots ,v_M) = \langle F_{b_1}(v_1),\dots | \mathcal{P} | F_{a_1}(u_1),\dots \rangle \,, \la{mainP} 
\eeq
involving any number of gluons and bound states in both incoming and outgoing states. As a result of the $\mathbb{Z}_2$ symmetry of the pentagon $\mathcal{P}$
\beq
\mu_a=\mu_{-a}\qquad  \text{and}  \qquad P_{\vec{a} | \vec{b}}=P_{-\vec{a} |- \vec{b}} \, ,
\eeq
such that in practice only the overall sign of the helicities in~(\ref{mainP}) matters. Later we will introduce another class of transitions (the so called charged transitions) for which this will not be the case.

The goal of this section is to present the form of the most general multi-particle pentagon transition involving the lightest gluonic excitations alone. Bound states can be understood by fusing these excitations together and will be the subject of the next section.

\subsection{Elementary Transitions}

We start by recalling what is known about these transitions from our analysis of the one and two particle OPE contributions \cite{short,data,2pt}. In \cite{short} we bootstrapped the direct transitions involving a single gluon in both the bottom and top of the pentagon. In pictures, 
\beq \la{singleGluonTransition}
\begin{array}{l}
\def\svgwidth{14cm}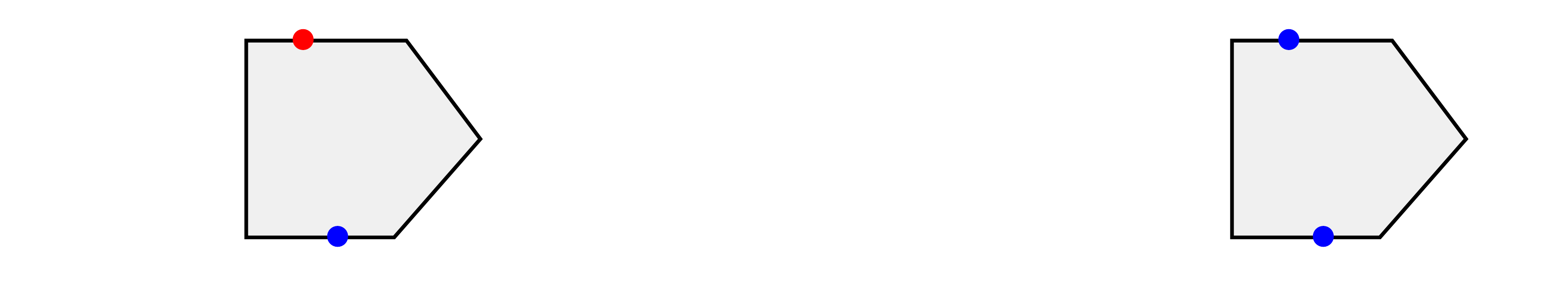 
\end{array}
\eeq
where blue/red dots represent insertions of a positive/negative helicity gluon field. Note that in our conventions $P_{1|1}$ corresponds to inserting the field $F$ on the bottom and its conjugate $\bar F$ on the top. It is thus the transition that preserves the helicity of the excitation which is flowing in between, whereas $P_{1|-1}$ violates it.%
\footnote{These two transitions were respectively denoted as $P(u|v)=P_{FF}(u|v)$ and $ \bar P(u|v)=P_{F\bar F}(u|v)$ in~\cite{data}.}

The so called direct transitions (\ref{singleGluonTransition}) obey a set of axioms, which was proposed and used in \cite{short} to bootstrap their finite coupling expressions. One of these postulates relates these transitions to the flux-tube S-matrices $S_{a,b}$ for the gluonic excitations. It was dubbed the \textit{fundamental relation} in \cite{short} and simply reads
\beqa
P_{a|b}(u|v)=P_{b|a}(v|u) S_{a,b}(u,v) \,. \la{fund}
\eeqa
Ironically, the fundamental relation is the most powerful of all axioms and yet the least well understood. 

The helicity preserving and the helicity violating transitions are not independent. Instead, one can relate them by using the so called mirror transformation $u\to u^\gamma$ which allows us to move particles from one edge to its left neighbour. (Similarly, we can use the inverse transformation $u\to u^{-\gamma}$ to move a particle to the right.) The point is that under one such move the gluon $F$ changes into a $\bar F$ and vice-versa \cite{short,data}. In this way we can relate the two transitions in (\ref{singleGluonTransition}) through the mirror axiom, 
\beq
P_{a|b}(u^{-\gamma}|v)= P_{-b|a}(v|u) \,, \la{mirrorMove}
\eeq
as depicted in figure \ref{mirror}.a. 

\begin{figure}
\centering
\def\svgwidth{16cm}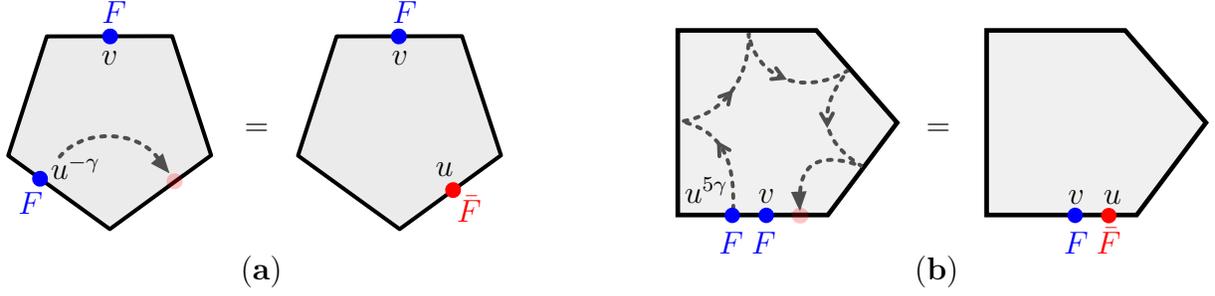
\caption{(\textbf{a}) The inverse mirror transformation $u\rightarrow u^{-\gamma}$ sends an excitation to the neighbouring edge on the right and simultaneously flips its $U(1)$ charge. (\textbf{b}) A sequence of five mirror rotations sends the excitation all the way around the pentagon.}
\la{mirror}
\end{figure}

The mirror transformations can be composed to move particles around the pentagon from one edge to \textit{any} other edge. In this regard, one can easily convince oneself that the above relations, combined with the identity 
\beq
P_{-a|b}(u^{2\gamma}|v)= \frac{1}{P_{a|b}(u|v)} \,, \la{2gamma}
\eeq
(which was discussed in great detail in appendix B.1 of \cite{2pt},) suffice to perform any sequence of mirror transformations. 
The simplest application of such manipulations was presented in~\cite{2pt} where the creation (annihilation) amplitudes for two gluons were derived from the direct transitions~(\ref{singleGluonTransition}) by bringing both particles to the top (bottom). For illustration, using~(\ref{2gamma}) to move the particle $u$ from bottom to top leads to the creation amplitude $P_{a,b}(0|u,v)={1}/{P_{a|b}(u|v)}$.
Since a creation amplitude can also be regarded as an annihilation amplitude, the same relation can also be cast as 
\beq
P_{a,b}(u,v|0)=\frac{1}{P_{b|a}(v|u)} \,. \la{creationDirect}
\eeq
We note, in particular, that an immediate consequence of (\ref{fund}) is the relation\footnote{Unitarity for the S-matrix yields $S_{a,b}(u,v)=1/S_{b,a}(v,u)$ and has been used to arrive at~(\ref{Watson}).} 
\beq
P_{a,b}(u,v|0)=S_{a,b}(u,v) P_{b,a}(v,u|0)  \,. \la{Watson} 
\eeq
It is reassuring for the consistency of the full bootstrap program to see this relation coming out. The point is that while the physical origin of (\ref{fund}) is still elusive, the relation~(\ref{Watson}) is the celebrated Watson equation \cite{Watson}. It translates the simple statement that once two incoming (or outgoing) particles are swapped one should pay their corresponding S-matrix. Intriguingly, for us the Watson equation is a consequence of the (more mysterious) fundamental relation where one performs the very unorthodox manipulation of swapping an incoming with an outgoing particle. 

We end this section by noting that we can compose several of the above moves to check many other identities with a clear geometrical interpretation. If we rotate all particles in the pentagon towards their neighbouring edge, for instance, the pentagon transitions better be left invariant. Indeed, 
\beq
P_{a|b}(u^{-\gamma}|v^{-\gamma})=P_{-a|-b}(u|v) = P_{a|b}(u|v) \la{shift}
\eeq
follows immediately by applying (\ref{mirrorMove}) twice. A slightly more interesting relation is 
\beq
P_{a,b}(u^{5\gamma},v|0)=P_{b,-a}(v,u|0)  \,. \la{monodromy}
\eeq
It states that if we take an annihilation form factor with two particles in the bottom of the pentagon and carry clock-wise the leftmost particle all around the pentagon, through a sequence of five mirror rotations, we end up with the annihilation amplitude where this particle is now standing on the right with its helicity flipped, see figure \ref{mirror}.b. Again, using the relations given before, it is straightforward to establish this relation.\footnote{For the impatient reader, one sequence that does the job is
\beq
\,\,\,\,P_{a,b}(u^{5\gamma},v|0)
\!\stackrel{(\ref{creationDirect})}{=}\!\frac{1}{P_{b|a}(v|u^{5\gamma})}
\!\stackrel{(\ref{shift})}{=}\!\frac{1}{P_{b|a}(v^{-\gamma}|u^{4\gamma})} 
\!\stackrel{(\ref{mirrorMove})}{=}\!\frac{1}{P_{-a|b}(u^{4\gamma}|v)}  
\!\stackrel{(\ref{2gamma})}{=}\! P_{a|b}(u^{2\gamma}|v)  
\!\stackrel{(\ref{2gamma})}{=}\! \frac{1}{P_{-a|b}(u|v) }
\!\stackrel{(\ref{creationDirect})}{=} \!P_{b,-a}(v,u|0) \,.
\nn\eeq
}

\subsection{General Transitions}

With no loss of generality, we shall focus on the annihilation amplitudes $P_{a_1\dots a_N}(u_1,\dots, u_N|0)$ where all particles are incoming. After all, following the discussion of the previous section, we can easily move particles from bottom to top and vice-versa through a sequence of mirror transformations as 
\beq
P_{a_1,\dots, a_N |b_1,\dots, b_{M}} (u_1^{2\gamma},\dots,u_N|v_1,\dots ,v_M)=P_{a_2,\dots, a_N|-a_1,b_1,\dots, b_{M}} (u_2,\dots,u_N|u_1,v_1,\dots ,v_M) \,, \nn
\eeq
or
\beq
P_{a_1,\dots, a_N|b_1,\dots, b_{M}} (u_1,\dots,u_N^{-3\gamma}|v_1,\dots ,v_M) =P_{a_1,\dots a_{N-1}|b_1,\dots, b_{M},a_N} (u_1,\dots,u_{N-1}|v_1,\dots ,v_M,u_N) \nn \,.
\eeq
As usual, the strategy for determining these amplitudes will be to first postulate a set of axioms that these objects must satisfy and then look for its minimal (i.e.~simplest possible) realization. This set consists of the following three items:
\begin{enumerate}
\item {Watson Equation:}\\
\beq\la{WatsonA}
\def\svgwidth{15cm}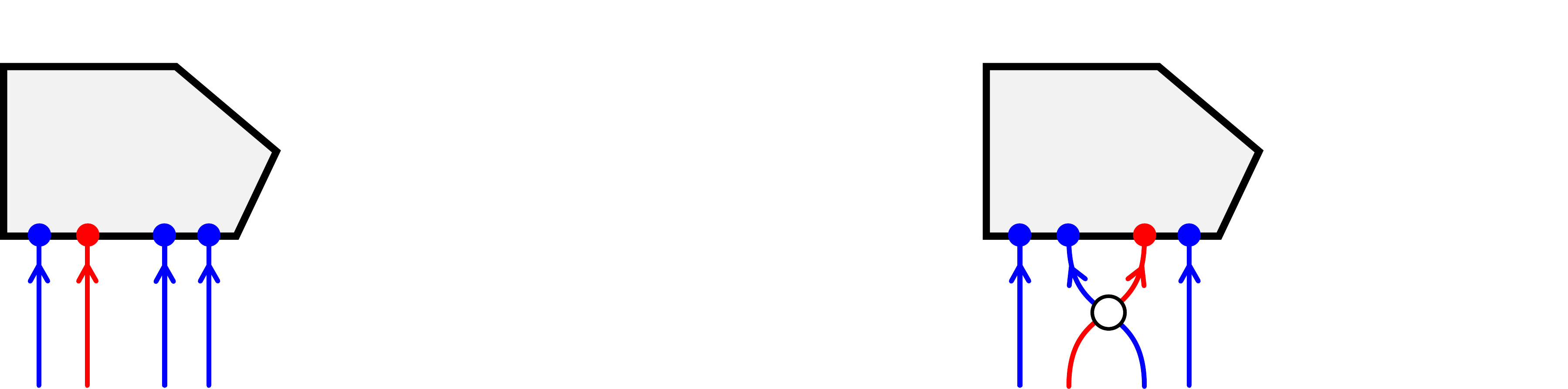
\eeq
\vspace{-1cm}
\item {Square Limit:}\\
\beq\la{SquareA}
\def\svgwidth{15cm}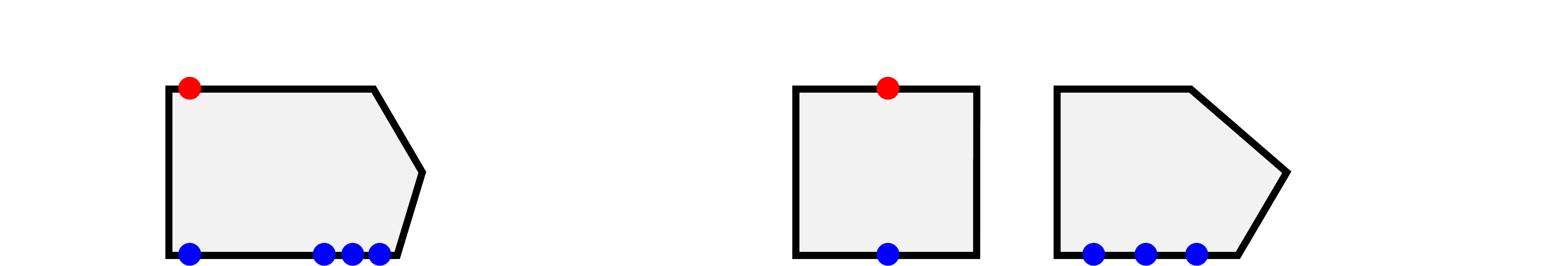
\eeq
\vspace{-.5cm}
\item {Monodromy:}
\beq\la{monodromyA}
\def\svgwidth{15cm}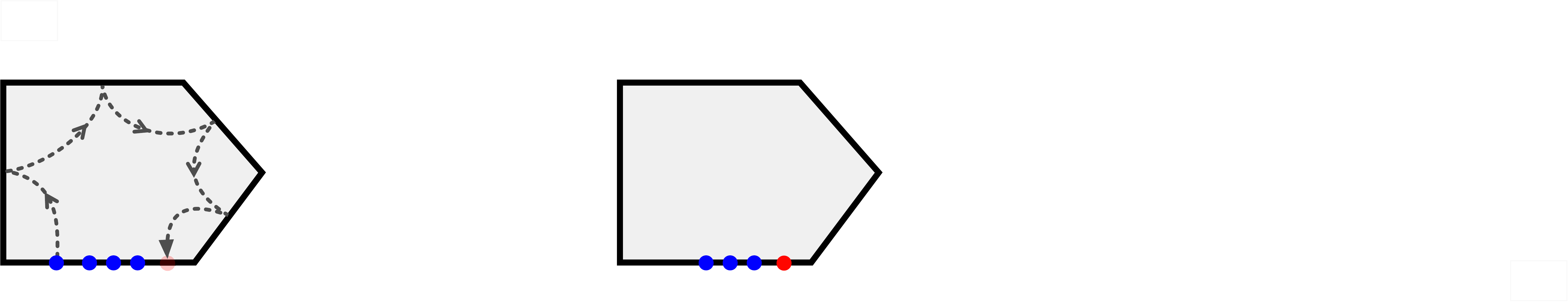
\eeq
\vspace{-.5cm}
\end{enumerate}
The first of these postulates was already discussed in the previous section and is the hallmark of any form factor analysis. The second one is less common. It states that if we first move the leftmost bottom particle to the top (with $u_1\to u_1^{2\gamma}$) and send both that particle as well as the (new) leftmost bottom particle towards the left edge of the pentagon then these two excitations will perceive a square geometry. This is because the right cusp together with all the other excitations effectively becomes infinitely far away from these two excitations~\cite{short,data}. Hence the result factorizes. In momentum space, this limit extracts the residue at $u_1=u_2$ and results in a simple factorized square measure $\mu_a(u_1)$ related to their direct transition through\footnote{In the past we sometimes used $\mu(u)$ or $\mu_F(u)$ to denote the fundamental gluon measure $\mu_1(u)$ and $\mu_{D^nF}(u)$ for the bound-state measure $\mu_{n+1}(u)$.}
\beqa
 i  \,\underset{u=v}{\operatorname{{\rm residue}}}  \,P_{a|b}(u|v)= \frac{\delta_{a,b}}{\mu_a(u)} \,, \la{measureDef}
\eeqa
see \cite{data}. The last axiom (\ref{monodromyA}) is even simpler to digest. It states that if a (leftmost) particle goes around the pentagon clockwise then after five mirror moves it ends up back at the original edge (but now at the rightmost position and with a flipped helicity). (We saw how this axiom was satisfied by the two particle annihilation form factor at the end of the previous section.)

The goal now is to come up with an ansatz for solving these axioms. The simplest thing to try is a totally factorized ansatz. Based on the two-particle examples discussed in the previous section, the simplest possible guess would be
\beq
P_{a_1,\dots,a_N}(u_1,\dots,u_N|0)= \prod\limits_{i<j} P_{a_i,a_j}(u_{i},u_{j}|0)= \frac{1}{\prod\limits_{i>j}P_{a_i |a_j}(u_{i}|u_{j})}\, , \la{factorizedCreation}
\eeq
where in the last equality we used the explicit form of the two-particle form factor (\ref{creationDirect}). 
Remarkably, this simple guess goes through all the pentagon axioms and, we conjecture, plays the role of the minimal solution we were looking for. It is quite elementary to check all three axioms. Both the Watson relation (\ref{WatsonA}) and the monodromy condition (\ref{monodromy}) follows immediately from the Watson equation (\ref{Watson}) and the monodromy condition (\ref{monodromy}) for the two-particle form factor.
The square limit axiom (\ref{SquareA}) follows trivially once we use (\ref{measureDef}) and~(\ref{2gamma}).

To finish our task, it remains to construct the most general transition by starting with the annihilation form factor (\ref{factorizedCreation}) and moving particles around as described before. Given that the algebra is straightforward, we merely quote the result here and let the more diligent readers work out the details. We find 
\beq
P_{a_1,\dots, a_N |b_1,\dots, b_M} (u_1,\dots,u_N|v_1,\dots ,v_M) = \frac{\prod\limits_{i,j}P_{a_i|b_j}(u_{i}|v_{j})}{\prod\limits_{i>j}P_{a_i |a_j}(u_{i}|u_{j})\prod\limits_{i<j}P_{b_i| b_j}(v_{i}|v_{j})}\, , \la{full}
\eeq
which could hardly be any simpler. This factorized result had been anticipated in \cite{short}. At leading order in perturbation theory (and for identical particles) it was recently confirmed in~\cite{Belitsky:2014rba}.

Since the fundamental transitions $P_{1|1}$ and $P_{1|-1}$ have been constructed at any coupling in \cite{short}, our conjecture (\ref{full}) provides a full finite coupling solution for the general pentagon transition involving any number of gluons. Still, this does not exhaust the full gluonic sector since we did not include bound states in the game yet. What we shall find is that the expression (\ref{full}) with the naive enlargement $a_i,b_i\in \mathbb{Z}$ suitably extends the pentagon transitions to include bound states.  The purpose of the next section is to establish it and, more importantly, to construct the fundamental transitions $P_{a|b}(u|v)$ for the bound states.

Finally, let us stress again that the problem we are solving here is very similar to the same sort of multi-particle bootstrap equations that arise in the computation of form factors in integrable models. (In some limits, our pentagon transitions fall precisely into a class of form factors previously considered in integrable theories; see e.g. \cite{O6paper} where we identified the scalar pentagon  transitions at strong coupling with the form factors of so called branch-point twist fields that recently arose in the study of entanglement entropy in integrable theories in \cite{CC-AD}.)
Yet, such a simple factorized ansatz as in (\ref{full}) for the multi-particle form factors --  constructed trivially out of the simplest possible form factors -- is \textit{not} the norm. What we are finding here is \textit{way} simpler than usual in this respect. Mathematically, it is the double mirror move (\ref{2gamma}) that is underlying this `miracle'. Thanks to it, it became possible (and in fact straightforward) to construct multi-particle form factors in terms of the single-particle transitions directly. Interestingly, this double mirror move (\ref{2gamma}) is not disconnected from the fundamental relation (\ref{fund}). In fact, it is clear that the Watson equations (\ref{Watson}) together with the double mirror move (\ref{2gamma}) \textit{imply} the fundamental relation. In other words, understanding this double move should shed light on the origin of \textit{both} the mysterious fundamental relation and the multi-particle ansatz.

\subsection{Charged Transitions} \la{ChargedSection}
\begin{figure}[t]
\centering
\def\svgwidth{11cm}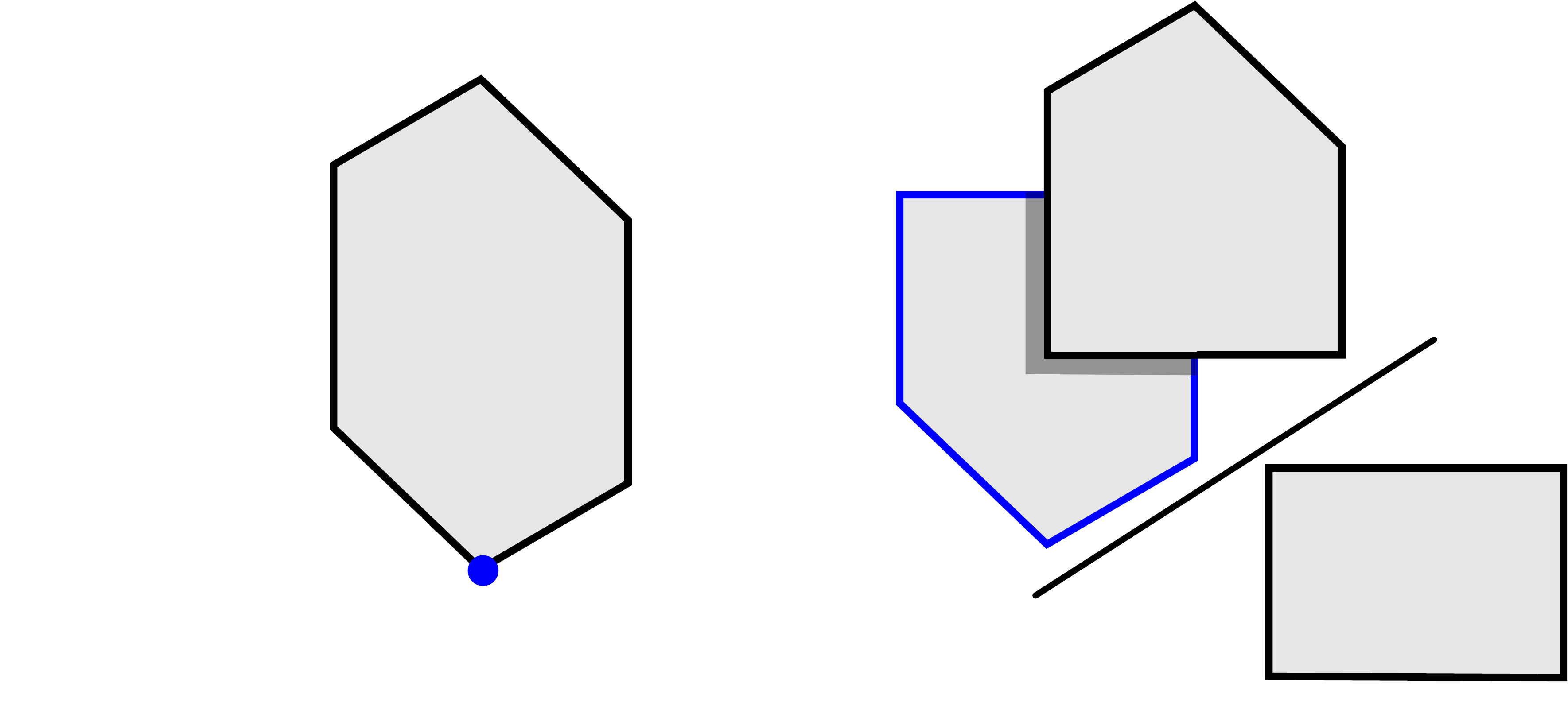
\caption{In the notations of~\cite{data}, the hexagon NMHV component $\cW^{(1111)}$ corresponds to charging the bottom edge (here numbered 1) with four $\eta$'s. At tree level, this component contains the insertion of a gauge field $\Blue{F}$ at the bottom cusp \cite{superloopskinner,superloopsimon}. In the OPE decomposition, charging the bottom edge this way amounts to replacing the bottom creation pentagon transition $P(0|\psi)$ with the charged transition $P^*(0|\psi)$.}
\la{HexagonNMHV}
\end{figure}
What we described so far were pentagon transitions entering the analysis of bosonic Wilson loops dual to MHV amplitudes.
The N$^k$MHV amplitudes are dual to super Wilson loops~\cite{superloopskinner,superloopsimon} which differ from their bosonic counterparts by additional insertions of adjoint fields at their edges and cusps. To address these more complicated objects within the OPE, we need to generalize the pentagon transitions to charged (or super) transitions. The simplest such charged transitions roughly correspond to inserting a gauge field $F$ on the pentagon. They allow us to describe NMHV components as the one depicted in figure \ref{HexagonNMHV}. 
The claim \cite{data} is that one can obtain these NMHV components by simply replacing the bottom transition $P_{a_1,\dots}(0|u_1,\dots)$ of an MHV amplitude with its charged counterpart $P^*_{a_1,\dots}(0|u_1,\dots)$. Here we bootstrap such charged transitions.

\begin{figure}[t]
\centering
\def\svgwidth{14cm}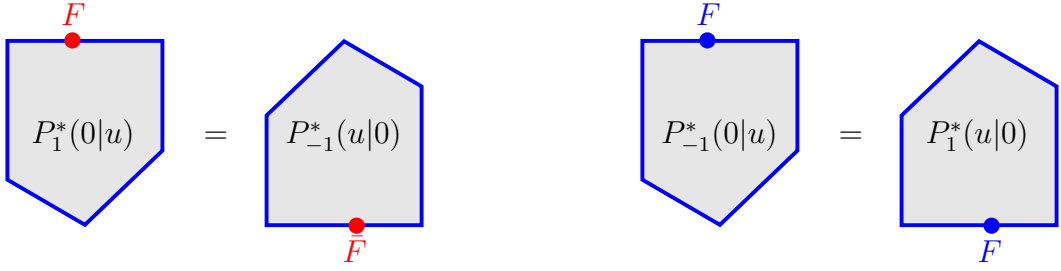
\caption{The four possible charged transitions for the creation or annihilation of a single gluon excitation, $\Blue{F}$ or $\Red{\bar F}$. The equalities in the figure follow from the rotation symmetry of the pentagon.}
\la{chargedtrans}
\end{figure}

The first case of interest is the form factor for creating a single gluon in a charged pentagon, see figure \ref{chargedtrans}. This was studied in \cite{data} and argued to be given by
\beq
P^*_{a} (0|u) =P^*_{-a} (u|0)=h_a(u) \,,
\eeq
where\footnote{Note that this form factor is even~\cite{data}, i.e.~$P^*_{a} (-u|0)=P^*_{a} (u|0)$.}
\beq
h_{\pm 1}(u) =\[ \frac{x(u+i/2)x(u-i/2)}{g^2} \]^{\pm 1} \,, \la{h1}
\eeq
in terms of the Zhukowsky variables 
\beq
2\, x(u)=u+\sqrt{u^2-4g^2} \,. \la{zu}
\eeq
We see that this form factor breaks the $\mathbb{Z}_2$ symmetry between the gluons $F$ and $\bar F$, as expected. (In particular, given that we are charging the pentagon with an $F$, it is natural that the creation of an $F$ is enhanced at weak coupling in comparison with the bosonic form factor $P_{a} (0|u)=1$, while the creation of an $\bar F$ is suppressed.)
More generally, we expect all the charged transitions to break this symmetry but still respect the very same axioms as written above. The simplest way of accommodating for such a thing is to adopt the factorized ansatz
\beq
P^*_{a_1,\dots |b_1,\dots} (u_1,\dots |v_1,\dots ) = P_{a_1,\dots |b_1,\dots} (u_1,\dots |v_1,\dots ) \prod_{i} h_{-a_i}(u_i) \prod_{i} h_{b_i}(v_i) \,. \la{Pcharged}
\eeq
One can easily check that the three axioms (\ref{WatsonA}), (\ref{SquareA}) and (\ref{monodromyA}) are satisfied for these charged transitions. The Watson equation~(\ref{WatsonA}), for instance, clearly continues to work, since the new addition is symmetric. The remaining two axioms rely on two simple properties of $h_a(u)$ \cite{data}:
\beq
h_a(u^{\gamma})=h_{-a}(u) \,, \qquad h_{-a}(u) h_a(u) =1 \,.\la{id23}
\eeq 
The first relation reflects the fact that a gluon swaps its helicity under a mirror move and it ensures that the monodromy relation (\ref{monodromyA}) is also observed for the charged transitions. Because of the second relation, we see that if we set two rapidities in the bottom and top to be the same (for gluons of the same kind) then the corresponding $h$'s in the new factor cancel out. As such, the square limit axiom (\ref{SquareA}) also continues to hold. (Conversely, we could have used the square limit axiom to further motivate the second identity in (\ref{id23}) which was adopted as an axiom in \cite{data} when bootstrapping the form factor $h_a(u)$.)

\section{Bound States and Fusion} \la{section3}

The lightest gluons studied thus far are only the tip of the gluonic sector. The latter also comprises heavier excitations which are bound states of the twist-one gluons. In this section we explain how the transitions for these bound states can be obtained from the ones for their constituents. 

\begin{figure}[t]
\centering
\def\svgwidth{10cm}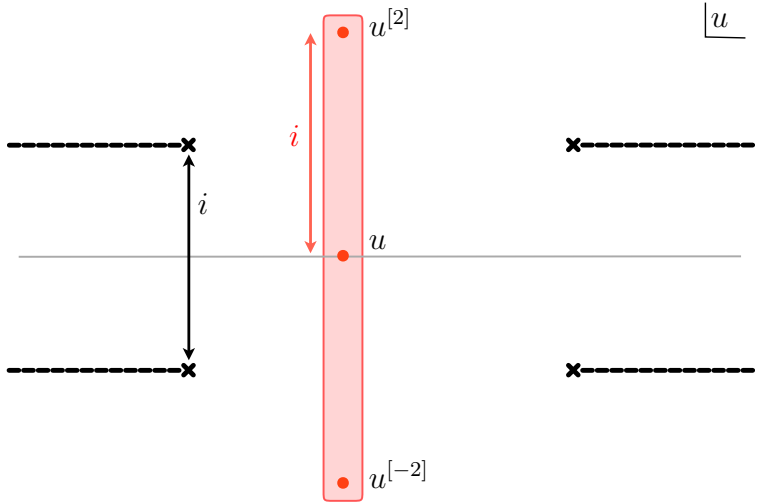
\caption{A bound state of $n$ gluons can be described by a vertical string of $n$ Bethe rapidities separated by $i$. The right prescription is to build the string for a centre-of-mass rapidity~$u$ lying within the strip $-2g<\text{Re}\,(u)<2g$. This (typically) means that the string is sitting in-between the branch points present in the complex rapidity plane of a (twist-one) gluon. This is what is shown here with the crosses representing the branch points at $\pm 2g \pm i/2$ and the dashed lines the outward cuts connecting them.}\label{Cuts}
\end{figure}

\subsection{Fusing the Transitions}\la{fusion-trans}

A bound state is no more than the collection of its constituents, each carrying (typically) a complex momentum. In integrable models, this description often becomes extremely simple when written in rapidity space. This is the case here, and, in these terms, a bound state of $n$ gluons is just a so called Bethe string of~$n$ rapidities, with any two close-by rapidities being separated by $i$ from each other. Accordingly, the energy and momentum of the bound state are obtained by summing over its fused elements,
\beq
E_{a}(u) = \sum\limits_{k=1}^{|a|}E_1(u^{[+2k-|a|-1]}) \,, \qquad p_{a}(u) = \sum\limits_{k=1}^{|a|}p_1(u^{[+2k-|a|-1]})  \,, \la{EP}
\eeq
with $u^{[j]} = u+i j/2$ and $u =\tfrac{1}{|a|} \sum\limits_{k=1}^{|a|}u^{[+2k-|a|-1]}$ the center-of-mass rapidity.

\begin{figure}[t]
\centering
\includegraphics[scale=.45,  trim= 0 70 0 100 ]{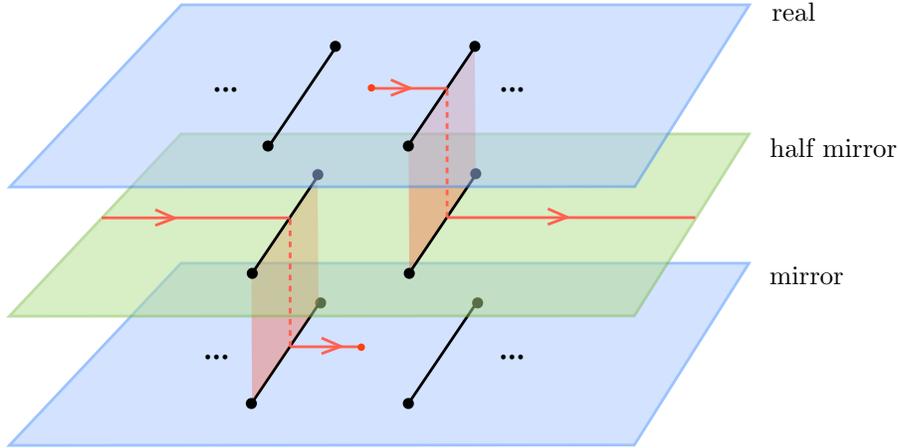}
\caption{Riemann surface of the gluonic excitation (with imaginary part of $u$ growing to the right). The mirror path $\gamma$ from the physical (or real) to mirror kinematics is depicted in red. It consists of a sequence of two similar steps that each can be thought of as an half-mirror rotation. Starting from the real sheet, we can access the half-mirror (or Goldstone) sheet by crossing the first Zhukowsky cut in the upper half rapidity plane. (In the physical sheet there are infinitely many other cuts in the lower/upper half planes, as indicated by the black dots.) This half mirror sheet contains the Goldstone point $E = -ip = |a|$, whose presence follows from symmetry considerations~\cite{AldayMaldacena}. It is realized at $u=\infty$ which is a regular point on this sheet. Yet another nice feature of this sheet is that it only contains finitely many cuts. There are only two of them for the twist-one gluon and $|a|+1$ for the bound state $F_a$. (Only the two outermost cuts are depicted on the green sheet, with the $|a|-1$ remaining ones implicitly lying in-between them.) When sitting on this sheet, we are half way to both real and mirror kinematics. To complete our trip we should cross the bottom-most cut in the lower half plane of the half mirror sheet. This brings us to the mirror sheet whose analytical properties are identical to those of the real one, but where $E$ and $p$ have exchanged their role.}\label{MirrorF}
\end{figure}

We must add that there is an important caveat here. As functions of $u$, the energy and momentum of a single gluon, i.e.~$E_1(u)$ and $p_1(u)$, both have a rich cut structure. It is therefore not enough to write~(\ref{EP}); we also need to specify where in this complicated Riemann surface should the string be formed. The prescription turns out to be rather simple (see section 4.1 of \cite{BenDispPaper}): we should construct the string inside the strip $-2g<\text{Re}\, (u)<2g$, as illustrated in figure~\ref{Cuts}. Practically, this is most easily done after moving the string upwards (that is, towards large positive imaginary values) such that all its constituent rapidities lie far away from all the cuts.   In other words, we can safely fuse gluons into bound states by first going to the so called \textit{half mirror sheet} -- represented by the middle (green) sheet in figure~\ref{MirrorF}. Once done, we can analytically continue the outcome back to wherever we want, and, in particular, to the original \textit{real sheet} -- that is, to the top (blue) sheet in the same figure. 

These bound states can also be thought of as fundamental particles, not any differently from the twist-one gluons. Like the latter excitations, they admit a mirror transformation that flips their energy and momentum. It is implemented by carrying the centre-of-mass rapidity $u$ of the string through the path ${\gamma}$ represented in figure~\ref{MirrorF}. The very same transformation implements the mirror rotation for the lightest gluons. Therefore, as we take the string through this path, all its (light) constituents cross the cuts and undergo a mirror transformation, resulting in
\beq
E_a(u^\gamma)=i p_a(u) \,, \qquad p_a(u^\gamma)=i E_a(u)\, ,
\eeq
regardless of how big the $a$-string is.

We can also study more complicated observables, like the S-matrix $S_{a,b}$ between two bound states. Following the familiar procedure, it can be obtained by fusing the S-matrices between the strings elements, yielding
\beq\label{Sab-fusion}
S_{a,b}(u,v) = \prod\limits_{k=1}^{|a|}\prod\limits_{j=1}^{|b|}S_{1, \epsilon}(u^{[+2k-|a|-1]},v^{[+2j-|b|-1]}) \,, \qquad \epsilon=\text{sign}(ab) \,.
\eeq
So defined, the S-matrix is automatically unitary and crossing symmetric,
\beq
S_{a, b}(u, v)S_{b, a}(v, u) = 1\, , \qquad S_{a, b}(u^{2\gamma}, v)S_{a, -b}(u, v) = 1\, ,
\eeq
since both properties are fulfilled by the original $|a|=|b|=1$ (i.e.~twist one) S-matrix~\cite{short,data} and both immediately lift to the general case through~(\ref{Sab-fusion}).

It is now tempting to assume that the very same recipe work as well for the pentagon transitions. Namely, we are led to set that
\beq
P_{a|b}(u|v) = \prod\limits_{k=1}^{|a|}\prod\limits_{j=1}^{|b|}P_{1| \epsilon}(u^{[+2k-|a|-1]}|v^{[+2j-|b|-1]}) \,, \qquad \epsilon=\text{sign}(ab) \,, \la{Pab}
\eeq
for the general transitions among bound states of gluons. This is the main formula of this section: it links together the transitions for bound states and constituent gluons.

Further motivation for adopting the ansatz~(\ref{Pab}) is that it verifies all the defining axioms for $P_{a|b}$. Two of them are actually automatic. Namely, if properties~(\ref{fund}) and~(\ref{mirrorMove}) hold for the twist-one gluons, then~(\ref{Pab}) guarantees -- together with~(\ref{Sab-fusion}) -- that they carry over to the bound states. What is less apparent is that the representation~(\ref{Pab}) correctly embodies the square limit~(\ref{measureDef}).

Let us verify it in detail for the bound state of two gluons. In this case we have
\beq
P_{2|2}(u|v) = P_{1|1}(u^+|v^+)P_{1|1}(u^+|v^-)P_{1|1}(u^-|v^+)P_{1|1}(u^-|v^-) \,,
\eeq
with $u^{\pm} = u\pm i/2$ and similarly for $v$. According to~(\ref{measureDef}), this should have a simple pole at~$u=v$. Instead, it seems as if the right hand side had a double pole, since both the first and last transitions behave as 
\beq
P_{1|1}(u^{\pm}|v^{\pm}) \sim \frac{1}{i(u-v)} \frac{1}{\mu_1(u^{\pm})}\, ,
\eeq
when $u\sim v$. What saves the day is that the third transition, $P_{1|1}(u^{-}|v^{+})$, vanishes (linearly) when $u\to v$, while the remaining factor, $P_{1|1}(u^{+}|v^{-})$, happens to be regular. Both properties are manifest in the representation~(\ref{PabMirror}) given in appendix~\ref{first} and, in the end, guarantee that $P_{2|2}$ has the proper square limit. The associated square measure for the two-gluon~bound state, i.e.~$\mu_2(u)=  1/  (i\, \underset{u=v}{\operatorname{{\rm residue}}}  \,P_{2|2}(u|v))$, can therefore be written as%
\footnote{One can use that $P_{1|1}(u^{-}|v^{+})\sim -P_{1|1}(v^{-}|u^{+})$ for $u\sim v$, which follows from $P_{1|1}(u^{-}|v^{+})$ having a simple zero.}
\beq
\mu_2(u)=
 i  \,\underset{v=u}{\operatorname{{\rm residue}}} \,\frac{\mu_1(u^+)\mu_1(v^-)}{P_{1|1}(u^+|v^-)P_{1|1}(v^-|u^+)}\, ,  \, \la{wdf} 
\eeq
which perfectly agrees with the result anticipated in \cite{2pt}. (We recall that the argument given in \cite{2pt} in favour of (\ref{wdf}) was that one should be able to look for the bound-state measure as a pole of the two-gluon integrand for the hexagon.)

The algebra for the general case is essentially the same. In appendix \ref{ApB} we provide a detailed construction of the bound-state transitions and verify that they all have the proper behaviour in the square limit. As mentioned above, the subtle point in this construction is that it should be done in the right kinematical region, the most convenient of which being the half-mirror sheet. This is carried out in appendix \ref{ApB} together with the analytical continuation of the fused object back to the physical sheet. The summary of the final results together with a discussion of their weak coupling expansions (performed in the physical sheet) is presented in appendix \ref{weakA}.

Equipped with the direct transitions~(\ref{Pab}) it is not more complicated than before to construct the most generic multi-particle transition involving bound states. The multi-particle ansatz (\ref{full}), which we encountered above for the lightest gluons, but now with $a_i,b_i \in \mathbb{Z}$ perfectly does perfectly the job. We leave it as an instructive exercise to the reader to check that it obeys the defining axioms~(\ref{WatsonA}),(\ref{SquareA}) and~(\ref{monodromyA}). 

\subsection{Fusing the NMHV Form Factors} 

The same fusion procedure should apply for the charged transitions discussed in section~\ref{ChargedSection}. These transitions differ from the bosonic ones by a simple product of $h$'s, see (\ref{Pcharged}). Hence, to fuse them, one simply needs to fuse these $h$'s, hence obtaining $h_a(u)$ from $h_{\pm 1}(u)$. This is what is done in great detail below. The most interesting aspect of this exercise is that it illustrates neatly the importance of fusing in the right place (while avoiding most of the technicalities involved in the fusion of more complicated objects such as the pentagon transitions, see appendix \ref{ApB}). 

We start with the simplest bound state, $F_2(u)$, and its associated form factor
\beq
h_2(u)=h_1(u+i/2)h_1(u-i/2) \,.
\eeq
Using the expression (\ref{h1}) for $h_1(u)$ we immediately get 
\beq
h_2(u)=\frac{x^+(u+i/2)x^-(u+i/2)}{g^2} \,\frac{x^+(u-i/2)x^-(u-i/2)}{g^2}\, , \la{h2}
\eeq 
where $x^\pm(u)=x(u\pm i/2)$. Naively, the right hand side of this equation evaluates to $x(u+i)x(u)^2x(u-i)/g^4$, which is certainly correct for $|\text{Re}\,(u)|>2g$. However, this is not where we are instructed to fuse. Instead we should consider $|\text{Re}\,(u)|<2g$, such that for real $u$ we are right on top of the Zhukoswky cut in (\ref{zu}). To avoid this, it helps giving $u$ an infinitesimal imaginary part and write $x^-(u+i/2)=x(u+i0)$ and $x^+(u-i/2)=x(u-i0)$. The two options differ by the choice of branch in (\ref{zu}), 
\beq
2 \,x(u\pm i0)=u \pm i\sqrt{4g^2-u^2}\, ,
\eeq
and are simply the inverse of each other, i.e.~$x(u+i0)=g^2/x(u-i0)$. This more careful analysis of (\ref{h2}) yields then $g^2$ instead of $x(u)^2$ and eventually
\beq
h_2(u)=\frac{x(u+i)x(u-i)}{g^2} \,, \la{h2Result}
\eeq 
which is the right result. 

The algebra got tricky because we had to do the fusion in the vicinity of the cuts. As advocated earlier, this can actually be avoided by first going to the half-mirror sheet. We recall that to get there, starting from the real sheet for $-2g<u<2g$, it suffices to transport our bound state upwards, that is toward the green sheet depicted in figure \ref{MirrorF}. Since we cross the top Zhukowsky cut along the way, we observe that $x^- \to g^2/x^-$ while $x^+$ remains untouched. This gives 
\beq
h_1(\hat u)=\frac{x(u+i/2)}{x(u-i/2)} \,, 
\eeq
where the hat on $u$ reminds us that we are sitting in the half-mirror sheet. We can now safely fuse in this sheet wherever we want and in particular far from any cut. For example, for the positive helicity bound state $F_a(u)$, this immediately yields
\beq
h_a(\hat u)=\prod_{k=1}^{a} h_1({\hat u}^{[+2k-a-1]}) =\frac{x(u+ia/2)}{x(u-ia/2)} \, , \qquad a>0 \,.
\eeq
Finally, we can analytically continue the outcome back to the physical sheet. This means re-entering through the upper Zhukowsky cut, which is now found at $\textrm{Im}\,(u) = a/2$ and amounts to $x(u-ia/2) \to g^2/x(u-ia/2)$. This gives
\beq
h_a(u)=\frac{x(u+ia/2)x(u-ia/2)}{g^2} \,, \qquad a>0 \,,
\eeq 
which reproduces the particular case (\ref{h2Result}) for $a=2$. The generalization to negative helicity is straightforward. We conclude, therefore, that 
\beq
h_a(u)=\( \frac{x(u+ia/2)x(u-ia/2)}{g^2} \)^{\text{sign}(a)} \,, \la{ha}
\eeq 
which is the main result of this section.

The expression~(\ref{ha}) allows one to generalize the equation~(\ref{Pcharged}) to include bound states as well. It suffices to let the indices in this equation take values over all integers.

\section{Applications} \la{section4}

At this point we can collect all our findings and put them to good use in the study of gluon scattering amplitudes.

\subsection{MHV Hexagon at Finite Coupling}
\begin{figure}[t]
\centering
\includegraphics[scale=.5, trim= 0 70 0 100 ]{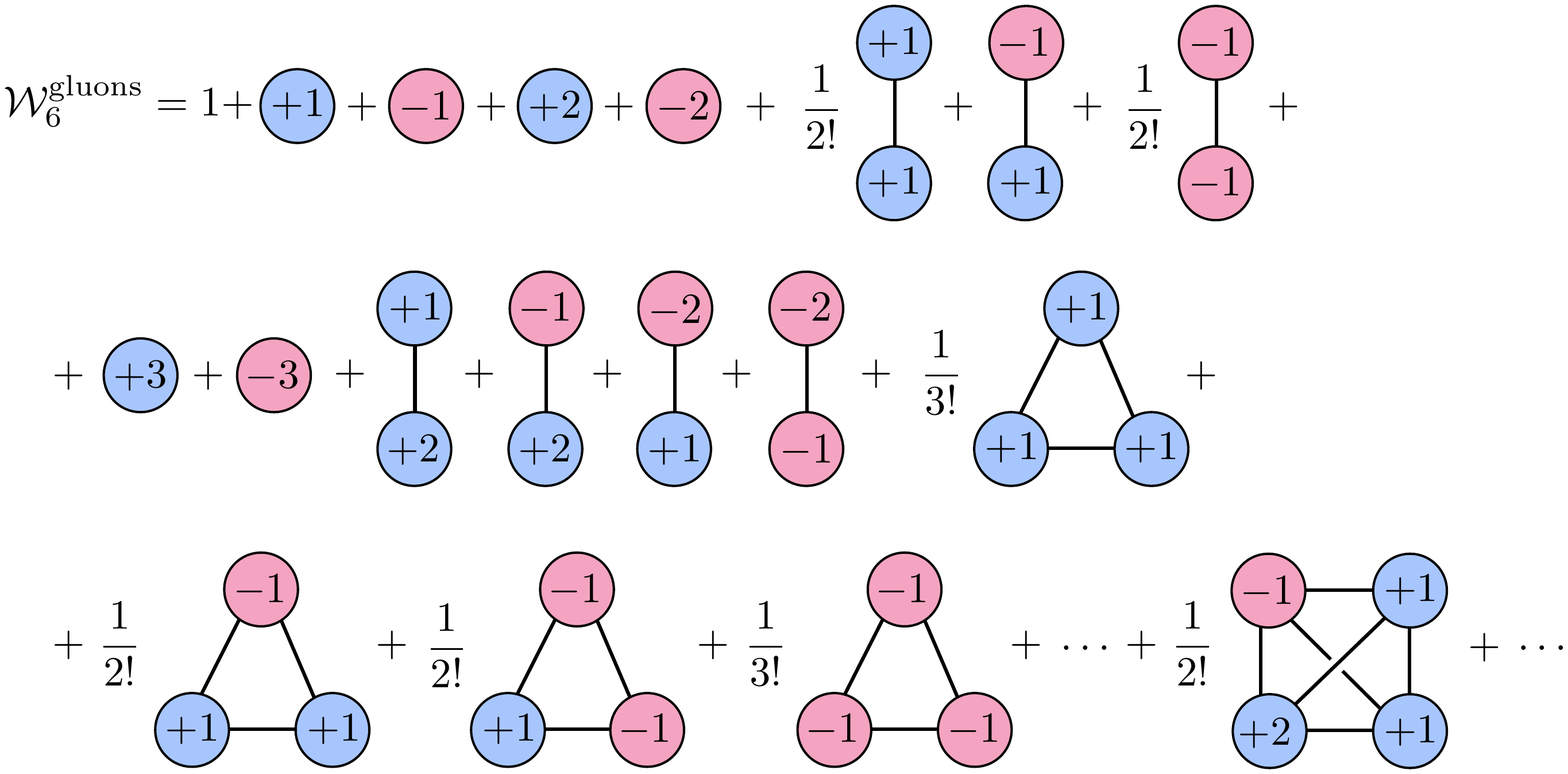}
\caption{Diagrammatic representation of the gluonic contribution to the hexagon OPE series~(\ref{W6sum}). A blob $a$ represents the effective measure $\hat{\mu}_a$ of a flux-tube excitation with $U(1)$ charge $a$. A  link connecting blob $a$ and $b$ stands for the kernel $1/P_{a|b}P_{b|a}$. Keeping only the blue/pink diagrams corresponds to projecting to the all positive/negative helicity subsector.}\label{example}
\end{figure}

We start by considering a bosonic hexagonal Wilson loop $W_6$ -- or rather its renormalized version $\mathcal{W}_6$ (see \cite{short}) -- which is relevant for MHV amplitudes. Putting together the results of the previous section for the creation and annihilation amplitudes of a generic gluonic state with $N$ excitations, we can readily write down the full gluonic contribution to the hexagon as 
\beq
\mathcal{W}_6^\text{gluons}=\sum_{N=0}^\infty \sum_{a_1\neq 0}\dots\sum_{a_N\neq 0} \Gamma\, \int  \frac{du_1\dots du_N}{(2\pi)^N}\, \frac{\hat \mu_{a_1}(u_1)\ldots \hat\mu_{a_N}(u_N)}{\prod\limits_{i\neq j} P_{a_i| a_j}(u_i|u_j)} \, ,
\la{W6sum}
\eeq
with the effective measure
\beq
\hat \mu_a(u)=\mu_a(u) e^{-E_a(u) \tau+i p_a(u) \sigma+i a \phi} \, .
\eeq
Here $\Gamma$ is a simple combinatorial factor, coming from the indistinguishability of the excitations, 
\beq
\Gamma=\prod_{k\neq 0} \frac{1}{n_k!} \, ,
\eeq
with $n_{k}$ the number of bound states with $U(1)$ charge $a_i=k$. 
It is worth noting that there is a nice pictorial representation that we can attach to each term in the sum (\ref{W6sum}). We can indeed represent each individual contribution as a fully connected graph with $N$ nodes, where each node stands for a measure $\hat \mu_a$ and each edge for a factor $1/P_{a|b}P_{b|a}$, see figure \ref{example}. Then $\Gamma$ is the usual symmetry factor of the corresponding graph. 

We stress that the result (\ref{W6sum}) is valid at any value of the coupling; each term in this sum being built out of the fundamental transitions and measures, which are all summarized in the appendix. Still, after so many conjectures -- both for the pentagon axioms as for their solutions -- it is essential to cross-check the integrability based prediction (\ref{W6sum}) against results obtained through more conventional methods.
With this goal in mind, we now turn our attention to the weak coupling analysis. 

\subsection{MHV Hexagon at Weak Coupling} \la{MHVhex}
To compare with perturbation theory, the zero-th order step is to expand all the ingredients in (\ref{W6sum}) at weak coupling. First we want to estimate how much they contribute. According to the expressions in the appendix we have
\beq
\hat \mu_a(u) = O(g^2) \,, \qquad P_{a|b}(u|v) = 
\left\{ \begin{array}{l} O(g^{-2})\,,\qquad  \text{sign}(ab)=+1\, , \\
\\
O(g^{0}) \,\,\,\,,\qquad  \text{sign}(ab)=-1\, . \end{array}\right.
\eeq
Therefore, a simple counting exercise shows that a multi-particle state with $N_+$ gluons with positive helicity and $N_-$ gluons of negative helicity contributes to the hexagon Wilson loop starting only at  
\beq 
l=N_+^2+N_-^2 \la{loops}
\eeq 
loops. We see that in practice, to compare with perturbation theory, we can safely truncate the sum (\ref{W6sum}) to a relatively small maximum number of particles. 

It is not totally straightforward, despite the truncation, to compare $\mathcal{W}_6^\text{gluons}$ in (\ref{W6sum}) with perturbative data. The reason is that conventional perturbative methods compute the \textit{full} Wilson loop, equivalently $\mathcal{W}_6$, and that it is typically challenging to isolate in it what comes purely from the gluonic excitations and what comes from the rest  (which includes notably excitations such as scalars and fermions). A state of two gluons with opposite helicity, for instance, has total twist two and zero total $U(1)$ charge and thus contributes in pretty much the same manner as a singlet pair of scalars or fermions \cite{2pt}. (Recall that in perturbation theory the contribution of a state with twist $n$ and total $U(1)$ charge $m$ scales as 
$
e^{-n \tau + i m \phi}
$
in the near collinear limit.)

There are fortunately two subsectors within (\ref{W6sum}) which we can match immediately and unambiguously against perturbation theory. They are obtained through restriction to multi gluons and bound states carrying helicities $a_i$ all of the same sign, being positive or negative. They collect, at weak coupling, contributions that have total twist equal (in magnitude) to the total helicity of the state, which, clearly, can only be coming from the gluons (see figure~\ref{truncation}). We denote these two collections as $\mathcal{W}_6^\text{gluons $+$}$ and $\mathcal{W}_6^\text{gluons $-$}$ respectively. 
Up to the first few terms, we have
\beqa
\mathcal{W}_6^\text{gluons $+$}&=&\int \frac{du}{2\pi} \hat \mu_1(u)+\int \frac{du}{2\pi} \hat \mu_2(u) +\frac{1}{2!}\int \frac{du dv}{(2\pi)^2} \frac{\hat \mu_1(u) \hat \mu_1(v)}{P_{1|1}(u|v)P_{1|1}(v|u)} \nn \\
&+&\int \frac{du}{2\pi} \hat \mu_3(u) +\int \frac{du \,dv}{(2\pi)^2} \frac{\hat \mu_1(u) \hat \mu_2(v)}{P_{1|2}(u|v)P_{2|1}(v|u)} \la{examples}\\
&+&\frac{1}{3!} \int \frac{du\, dv\, dw}{(2\pi)^3} \frac{\hat \mu_1(u) \hat \mu_1(v)\hat \mu_1(w)}{P_{1|1}(u|v)P_{1|1}(v|u)P_{1|1}(u|w)P_{1|1}(w|u)P_{1|1}(v|w)P_{1|1}(w|v)}+\dots\, , \nn
\eeqa
while the expression for $\mathcal{W}_6^\text{gluons $-$}$ is obtained by charge conjugation, or in other words by $\phi \to -\phi$.

We stress again that these two sums control all contributions that scale as $e^{-a \tau \pm i a \phi}$ when we expand the weak coupling results in the near collinear limit $\tau \gg 1$. 
More precisely, when expanding the perturbative result for $\mathcal{W}_6$ at large $\tau$ one finds
\beq
\mathcal{W}_6=1+\sum_{n=1}^\infty e^{-n \tau} 2\cos(n \phi) f_n(\sigma,\tau)+\sum_{n=2}^\infty e^{-n \tau} 2\cos((n-2) \phi) g_n(\sigma,\tau)+ \dots \,, \la{contamination}
\eeq
where, at $l$ loops, $f_n,g_n,\dots$ are polynomials of degree $l-1$ in $\tau$ and complicated (typically transcendental) functions of $\sigma$. What we can now easily predict -- to all loops -- is the full first sum
\beq
\sum_{n=1}^\infty e^{-n \tau\pm i n \phi} f_n(\sigma,\tau) = \mathcal{W}_6^\text{gluons $\pm$} \,.
\eeq

\begin{figure}[t]
\centering
\includegraphics[scale=1]{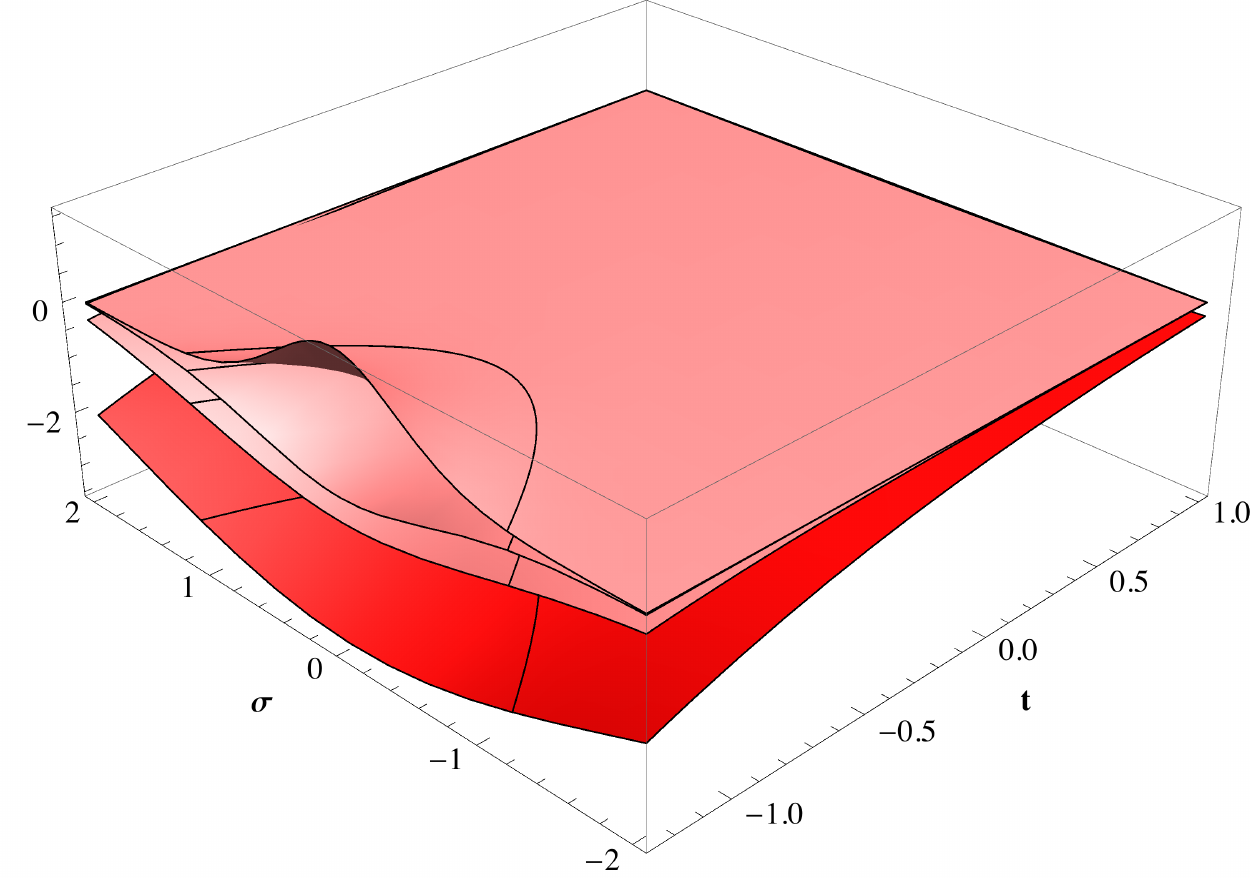}
\caption{The bottom curve (in red) is the exact result (\ref{scalingW}) plotted as a function of $\sigma$ and $t=\tau-i \phi$. The other three curves (in pink) from bottom to top are the differences between several truncations and the exact result (for $1$, $3$ and $6$ gluons respectively). For $t$ positive (not necessarily) large we see that all the truncations approximate the exact result perfectly. Only for (large) negative $t$, i.e.~very far from the near collinear limit, we do start noticing that we need to add more and more gluons to converge towards the exact result. The solid line indicates the OPE radius of convergence.  }\label{ScalingPlot}
\end{figure}

The easiest check is at one loop where the Wilson loop is simply given by the BDS ansatz \cite{BDS} and takes the simple form \cite{Straps}
\beqa
&&\mathcal{W}_6^\text{1-loop}=g^2(\pi^2/6-\text{Li}_2(1-u_1)-\text{Li}_2(1-u_3)-\log (u_1) \log ( u_3) \nn \\
&&\qquad \qquad \qquad \qquad +\,\,\text{Li}_2(u_2)+\log^2(1-u_2)-\log(1-u_2)\log({u_1}/{u_3}))\, , \la{ru1} 
\eeqa
where $u_i$ are the three cross-ratios of the hexagon. One could now expand it at large $\tau$, collect all terms that vanish as $e^{-n\tau + i n \phi}$ and compare their sum with $\mathcal{W}_6^\text{gluons +}$. (We could also compare  each of them individually with the corresponding term in $\mathcal{W}_6^\text{gluons +}$.) Alternatively, one can follow a shortcut and isolate $\mathcal{W}_6^\text{gluons +}$ from $\mathcal{W}_6$ by considering the double scaling limit where $-\tau + i \phi$ is held fixed with $\tau$ and $i\phi$ both very large \cite{Straps}. In this limit $u_2\to 0$ and $u_{1,3} \to \tilde u_{1,3}$ where 
\beq
\tilde u_1 =  \frac{1}{1+e^{-2\sigma}+e^{-\sigma-\tau+i \phi}} \,, \qquad \tilde u_3 =  \frac{1}{1+e^{2\sigma}+e^{\sigma-\tau+i \phi}} \,. \la{tildeu}
\eeq
In this limit we can thus drop the second line in (\ref{ru1}) and replace the cross-ratios in the first line by their tilded counterparts. All we have to do is now compare this with~(\ref{examples}) which, at this loop order, only receives contributions from single particle states. Using the explicit expressions in the appendix~\ref{lo-exp-app} for the measures $\mu_a$ we have
\beq
\mathcal{W}_6^\text{gluons $+$}=g^2 \sum_{a=1}^\infty e^{-a \tau+i a \phi}  \int \frac{du}{2\pi} \frac{ (-1)^a \Gamma(\frac{a}{2}+iu) \Gamma(\frac{a}{2}-iu)}{(\frac{a^2}{4}+
   u^2) \Gamma (a)}\, e^{2i u \sigma }   +O(g^4) \,.
\eeq
In perfect agreement with the perturbative data, these integrals can be computed and even resumed into 
\beq
\mathcal{W}_6^\text{gluons $+$}=g^2(\pi ^2/6-\text{Li}_2(1-\tilde{u}_1)-\text{Li}_2(1-\tilde{u}_3)-\log (\tilde{u}_1) \log (\tilde u_3))+O(g^4) \,, \la{scalingW}
\eeq
which is plotted in figure \ref{ScalingPlot} (after stripping off the overall power of $g^2$). This is the simplest check of our conjectures. We should stress that it already probes the leading order expressions for the measures of \textit{all} bound states.

It would be very interesting to push this comparative analysis to higher loops using~\cite{VerguPaper,DelDuca:2010zg,Remainder,Lance,Dixon:2013eka}. {Restricting to the maximal helicity sector would then amount to keeping all terms vanishing like $e^{-n\tau + i n \phi}$, up to powers of $\tau$, in the collinear limit.}%
\footnote{{Said differently, the double scaling limit~(\ref{tildeu}) only makes sense at higher loops up to powers of $\tau \sim -\frac{1}{2}\log{u_2}$, see~(\ref{OPEexp}) for illustration.}} Up to three loops, only single particle states contribute in this subsector, see counting (\ref{loops}). Therefore, one would merely need to correct the energy, momentum and measure of each bound state, using the expressions in the appendix~\ref{weakA}. With the technology developed in \cite{Papathanasiou:2013uoa, Papathanasiou:2014yva} it should be possible to compute each resulting bound-state integrals and, hopefully, resum them all.
At three loops, for instance, this should yield
\beq\la{OPEexp}
\[\mathcal{W}_6^{\text{gluons } +}\]_{3 \text{ loops}}=F_1(\tilde u_1,\tilde u_3)+\tau F_2(\tilde u_1,\tilde u_3)+ \tau^2 F_3(\tilde u_1,\tilde u_3)\, ,
\eeq
with $F_1,F_2,F_3$ in agreement with the three-loop data \cite{Lance,Dixon:2013eka}. Performing this single-particle exercise at higher loops is more academical since at four, nine, sixteen, $\dots$ loops we also need to include two, three, four, $\dots$ gluons in the OPE analysis.\footnote{These are given by multiple integrals which involve the multi-particle creation form factors and it would be fascinating to develop powerful techniques for taming them analytically.}
 Still, it might be of interest and hint at possible hidden structures similar to those found in the Regge limit in \cite{Pennington:2012zj} or unable a more direct connection with the multi-Regge limit along the lines of~\cite{Hatsuda:2014oza}.

{The tests that we have done were less thorough but easily extendable to any loop order. 
We simply compared the first few leading terms in the OPE with the near collinear expansion of the available perturbative data. More precisely, the leading two terms ($e^{-\tau+i \phi}$ and $e^{-2\tau+2i \phi}$) were already matched against the OPE to four loops in \cite{data,2pt,Dixon:2013eka,Dixon:2014voa}. In \cite{Dixon:2014voa} and~\cite{lancePrivate} one can find the predictions for the next three subleading terms ($e^{-3\tau+3i \phi}$, $e^{-4\tau+4i \phi}$ and $e^{-5\tau+5i \phi}$) to the same loop order. We also confirmed these against the OPE.\footnote{We thank Lance Dixon for sharing with us the expansion \cite{lancePrivate} of the four-loop amplitude to order $e^{-4\tau+4i \phi}$ and $e^{-5\tau+5i \phi}$ which made this comparison possible.} Given that these checks are already highly non-trivial we see them as sufficient evidence for our ans\"atze, but it would be definitely interesting to push these further. }

Finally, as already alluded to in the introduction, it is quite amazing that the information in these maximal helicity subsectors happens to be enough to bootstrap the hexagon Wilson loop up to four loops within the hexagon function program \cite{Dixon:2013eka,Dixon:2014voa}. It remains unclear to us why this is so and whether this will persist to higher loops.

\subsection{NMHV Hexagon}
Having presented our prediction for the full gluonic sector for the 6-point MHV amplitude we proceed to the 6-point NMHV case. As mentioned above, see figure \ref{HexagonNMHV}, we focus here on the component $\mathcal{W}^{(1111)}$ of the super loop which differs from the MHV amplitude by replacing the bottom creation form factor by its charged counterpart (\ref{Pcharged}). Accordingly, the NMHV integrand for the gluonic contributions follows from the MHV one (\ref{W6sum}) by the replacement
\beq
\hat{\mu}_a(u) \to \hat{\mu}_a(u) h_a(u) \,.
\eeq
It is remarkable that such a simple rule can accommodate for the difference between MHV and NMHV amplitudes. It is even more remarkable that these additional form factors are simply given by a bunch of Zhukowsky variables (\ref{ha}). Maybe this simplicity could find some interpretation in the context of the Q-bar equation approach \cite{Qbar}? Conversely, can this shed light on the physical origin of the Zhukowsky variables? 

Though the form factors appear as a minor modification, they have important effects. For instance, since they scale as $h_a = O(g^{-2\,\text{sign}(a)})$, they modify the loop counting~(\ref{loops}) to
\beq 
l=N_+^2-N_++N_-^2+N_-\, , \la{loopsNMHV}
\eeq 
which clearly favours positive helicity gluons as compared to negative helicity ones. Importantly, the \textit{all positive} helicity sector ($N_-=0$) is no longer simply related to the \textit{all negative} helicity one ($N_+=0$). The former starts at tree level while the latter shows up at two loops. As for the MHV analysis, these maximal helicity sectors are particularly interesting to consider in perturbation theory because they do not receive any sort of contaminations from non-gluonic excitations. We shall now analyze each of them at their leading order in perturbation theory following closely the discussion of the previous section. 

The all positive helicity sector $\mathcal{W}_6^{(1111)\,\text{gluons }+}$ starts at tree level and comes solely from $N_+=1$ gluons. At this order, the effect of the form factor (\ref{ha}) is merely to multiply the integrand by $\frac{1}{g^2}(u^2+\frac{a^2}{4})$. Clearly, this is equivalent to acting on the MHV result with a Laplacian with respect to $\sigma$ and $\phi$. More precisely,\footnote{It would be interesting to understand if there is any connection between (\ref{box1}) (or (\ref{reduction}) below) and the recent studies \cite{Henn:2013pwa} of various differential equations obeyed by Feynman integrals.}
\beq
\mathcal{W}_6^{(1111)\,\text{gluons }+}=\frac{ \Box}{g^2} \mathcal{W}_6^{\text{gluons }+} +O(g^2)\,, \qquad \Box =- \frac{1}{4} \( \partial_{\sigma}^2+\partial_\phi^2\) \,. \la{box1}
\eeq
Using our previous result for the positive-helicity contribution to the MHV amplitude~(\ref{scalingW}), we immediately get
\beq
\mathcal{W}_6^{(1111)\,\text{gluons }+}= (\tilde{u}_1+\tilde{u}_3-1) +O(g^2)\,. \la{tree1111}
\eeq
We see that acting with the Laplacian has decreased the degree of transcendentality such that the end result is rational, as expected for a tree-level amplitude. We also easily verify that~(\ref{tree1111}) vanishes in the collinear limit $\tilde{u}_1+\tilde{u}_{3}\rightarrow 1$. It is now straightforward to compare our prediction~(\ref{tree1111}) with tree-level NMHV amplitude. We just need to recall the existing relation between $\mathcal{W}^{(1111)}_6$ and the $(1111)$ component of the NMHV ratio function  $\mathcal{R}$, which reads
\beq
\mathcal{R}^{(1111)}_{6} =  \mathcal{W}^{(1111)}_6 / \mathcal{W}_{6} \,.\la{WR} 
\eeq
To leading order at weak coupling they are just the same,~$\mathcal{R}^{(1111)}_{6} =\mathcal{W}^{(1111)}_6+O(g^2)$.
Indeed, evaluating this component using the package~\cite{Bourjaily:2013mma} with the twistors given  in Appendix~A of~\cite{data} perfectly reproduces~(\ref{tree1111}) after taking the double scaling limit which isolates the positive helicity gluons (see discussion above equation~(\ref{tildeu})).

A somewhat similar strategy can be applied to computing the negative helicity contribution at weak coupling (which as explained earlier kicks in at two loops). Since the form factor for these gluons is the inverse of the above one, one can no longer simply use the MHV result. Instead what we expect now for 
\beq
\mathcal{W}_6^{(1111)\,\text{gluons }-} =g^4 \sum_{a=1}^\infty e^{-a \tau-i a \phi}  \int \frac{du}{2\pi} \frac{(-1)^a \Gamma(\frac{a}{2}+iu) \Gamma(\frac{a}{2}-iu)}{(\frac{a^2}{4}+
   u^2)^2 \Gamma (a)}\, e^{2i u \sigma }   +O(g^6) \,, \la{sumNMHV}
\eeq
is a transcendental weight four function which once acted upon by the Laplacian should reduce to the MHV result, 
\beq
\mathcal{W}_6^{\text{gluons }-} =\frac{\Box}{g^2} \mathcal{W}_6^{(1111)\,\text{gluons }-} +O(g^4)\,. \la{reduction}
\eeq
One could imagine evaluating each term in (\ref{sumNMHV}) and resumming the outcomes to compute this transcendental weight four function. A shortcut would be to extract it from the scaling limit of the two-loop super Wilson loop $\mathcal{W}_6^{(1111)}$ -- related to the two-loop ratio function \cite{Dixon:2011nj} through the simple relation (\ref{WR}) -- and then simply check (numerically for example) that it does resum~(\ref{sumNMHV}). It would be interesting to do this exercise. 

We performed a simpler check of our prediction (\ref{sumNMHV}) which nevertheless probes it almost entirely. By truncating the sum (\ref{sumNMHV}) and evaluating each integral by closing the contours in the lower-half plane we generate the double Taylor series at small $y\equiv e^{\sigma}$ and $x\equiv e^{-\tau-i \phi}$
\beqa
\mathcal{W}_6^{(1111)\,\text{gluons }-} &=&x \left(-\frac{\pi ^2 y}{6}-3 y-2 y \log ^2(y)+4 y \log (y)+\frac{y^3}{4}+O(y^5)\right) + \nn \\
&+& x^2 \left(\frac{\pi ^2 y^2}{24}-\frac{y^2}{16}+\frac{1}{2} y^2 
   \log ^2(y)-\frac{2 y^4}{9}+O(y^6)\right)+ \la{WNMHVexpansion}\\
   &+&x^3 \left(-\frac{\pi ^2 y^3}{54}+\frac{y^3}{54}-\frac{2}{9} y^3 \log
   ^2(y)-\frac{5}{27} y^3 \log (y)+O(y^5)\right)+O(x^4) \,. \nn 
\eeqa
If we now set all $\pi$'s to zero in this expansion we can compare all the rest with the double Taylor expansion of the two-loop \textit{symbol} of $\mathcal{W}_6^{(1111)} $ in the scaling limit using the recursive algorithm described in \cite{Dixon:2014voa}. We checked it to order $O(x^{30} y^{30})$ probing a total of 299 coefficients and finding a perfect match for all of them. (For the reader's convenience, we quote in the appendix~\ref{W1111Ap} the symbol of $\mathcal{W}_6^{(1111)} $ in the scaling limit.) 

This sort of analysis should be useful in constraining the NMHV ratio function at higher loops within the hexagon program framework. At three loops, for instance, we could verify the consistency between our OPE result and the bootstrapped NMHV amplitude~\cite{LanceMatt} at the level of the $O(e^{-\tau+i\phi})$ and $O(e^{-2\tau+2i\phi})$ term. The check of the latter contribution is especially interesting since it probes, for the first time, the loop corrections to the $N_+=2$ (i.e.~two gluons) integrand.%
\footnote{We thank Lance Dixon and Matt von Hippel for sharing with us their findings for the three-loop NMHV hexagon amplitude~\cite{LanceMatt} prior to publication.}
(For comparison, an analogous test at MHV level would require knowledge of the five-loop amplitude, which seems within the reach of the hexagon program~\cite{Dixon:2013eka,Dixon:2014voa} but is currently unavailable.)

\subsection{Heptagon}

The OPE series for the heptagon WL is significantly more bulky than for the hexagon. 
For the heptagon there are now two middle squares and we can have gluonic excitations with rapidities $\{u_1,\dots,u_N\}$ in the bottom square and $\{v_1,\dots,v_M\}$ in the top square. Three transitions now show up in the full sequence $\text{vacuum} \to \{u_i\}\to \{v_j\} \to \text{vacuum}$. Putting together (\ref{full}) and the measures for all the excitations we easily see that a term with $N_{\pm}$ ($M_{\pm}$) gluons of positive/negative helicity in the bottom (top) square shows up at 
\beq
l=N_{+}^2+ M_+^2 -N_+ M_+ + N_{-}^2+ M_-^2 -N_- M_- \la{loopsHep}
\eeq
loops. As for the hexagon, the contributions where the gluons in a given square all have the same helicity can be easily isolated in perturbation theory. For instance, if we have positive helicity gluons in both squares we get 
\beq
\mathcal{W}_{7}^{\textrm{gluons } +,+} =  \sum_{a, b \geqslant 1} e^{ia\phi_1+ib\phi_2}\int \frac{du\, dv}{(2\pi)^2} \mu_{a}(u)P_{a|b}(-u|v+i0)\mu_{b}(v) e^{ip_a\sigma_{1}+ip_b\sigma_2-E_{a}\tau_1-E_{b}\tau_2} + \ldots\, , \la{Wpp}
\eeq
where dots include disconnected terms (i.e.~transitions with vacuum at top and or bottom) as well as multi-particle transitions.\footnote{{The $i0$ prescription is such that the result will have a square limit, see \cite{data}.}} From (\ref{loopsHep}) we see that up to three loops only these single particle transitions ($N_+=M_+=1$) matter.\footnote{It is also interesting to note that the two-particle contribution, which first appears in the form $N_+ =2M_+= 2$ or $M_+ = 2N_+=2$, shows up at three loops for the heptagon while it kicks in at four loops for the hexagon. This indicates that multi-particle transitions can be more directly probed in perturbation theory using higher polygons.} We compared  this expansion with the one loop result along the lines of the discussion in section \ref{MHVhex} and found a perfect agreement. This is a nice check of all the bound-state (helicity preserving) transitions to leading order in perturbation theory.

There is actually a simpler check that probes the transitions more directly. We can consider the OPE sum for an heptagon with a charged bottom pentagon. This allows us to replace the first measures in (\ref{Wpp}) as $\mu_a(u) \to h_a(u) \mu_a(u)$ leaving $\mu_b(v)$ untouched, 
\beq\la{WppNMHV}
\def\svgwidth{15cm}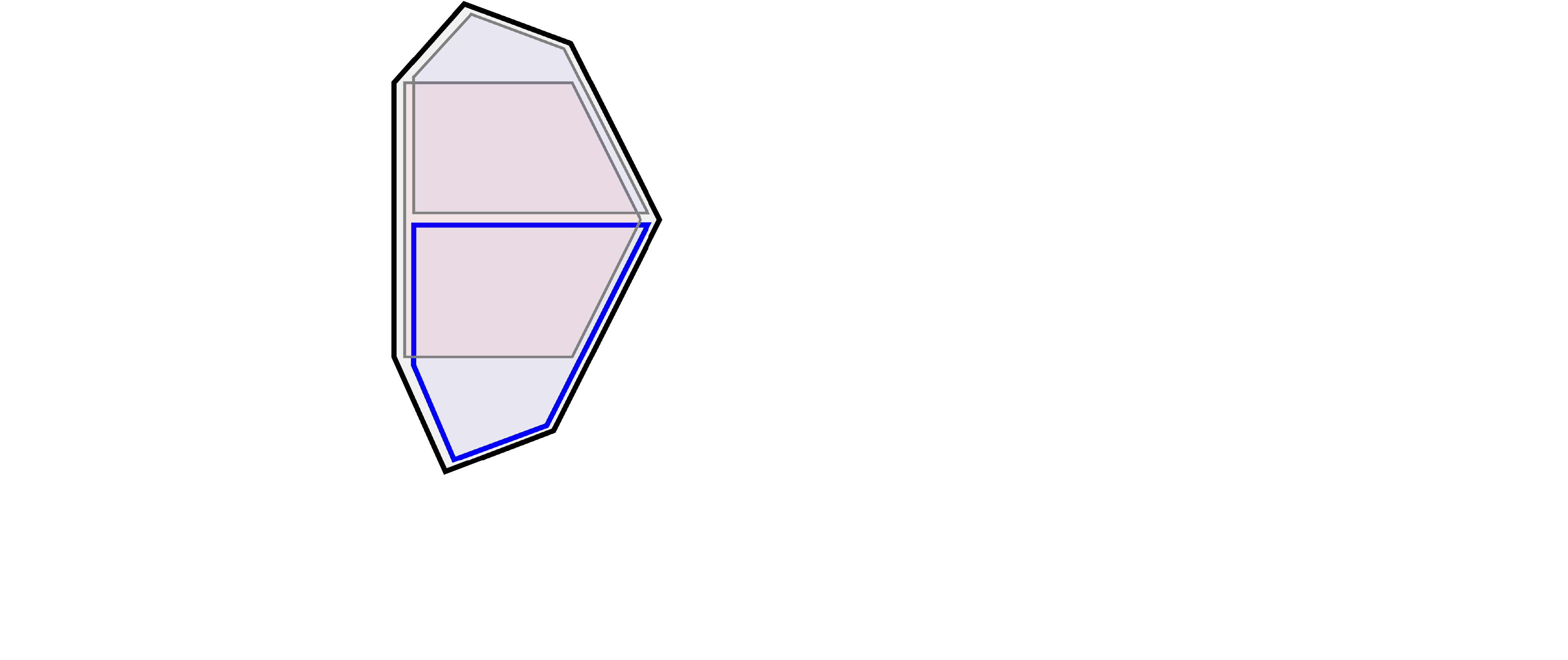
\eeq
Because $h_a=O(1/g^2)$ the resulting object will now start at tree level, that is one loop earlier than before. As discussed above and in \cite{data}, this generates an NMHV component with four $\eta$'s at the bottom edge. The corresponding ratio function can be straightforwardly extracted from the package \cite{Bourjaily:2013mma} by the command \\ \\ 
\verb"W1111=superComponent[{1, 2, 3, 4}, {}, {}, {}, {}, {}, {}]@treeAmp[7, 1]" \\ \\ 
We can then define the heptagon twistors \verb"Zs" to be as given in appendix A of \cite{data} and evaluate this component with the simple command \verb"evaluate@W1111". In the scaling limit where $\tau_j \to \infty$ with $\tau_j - i \phi_j$ fixed this gives
\beq\label{yy-eq}
-1+\frac{x_1 y_1}{x_1 y_1+x_1^2+1}+\frac{x_1 x_2 y_2 y_1}{\left(x_1 y_1+x_1^2+1\right) \left(x_2 x_1^2 y_2+x_2^2 x_1 y_1+x_2 x_1 y_1 y_2+x_2^2 x_1^2+x_1^2+x_2^2\right)} 
\eeq
where $x_j=e^{\sigma_j}$ and $y_j=e^{-\tau_j+i \phi_j}$. In this form the expression is amenable to a direct comparison with the OPE. When both $y_1,y_2\to 0$, for instance, only the first term in~(\ref{yy-eq}) remains; it stands for the vacuum contribution. As usual, we use it to normalize to our conventions the tree level amplitude, which means multiplying it by minus one. When $y_2\to 0$, with $y_1$ fixed, the second term survives. It corresponds to the contribution where we have the vacuum in the second square and is hidden in the dots in (\ref{WppNMHV}). Finally, and more interestingly, we have the last term which admits a double Taylor expansion in $y_1$ and $y_2$. The sum (\ref{WppNMHV}) resums into precisely this expression when we use the leading order transitions given in appendix \ref{lo-exp-app}.

We could as well consider mixed scaling limits, as $\tau_1-i \phi_1$ and $\tau_2+i \phi_2$ held fixed with $\tau_1,\tau_2$ very large. In this limit, we isolate once more the positive helicity gluons in the bottom square but also project into the negative helicity subsector in the top square. For MHV amplitudes, for instance, such contribution first shows up at two loops with a single gluon in each square, i.e.~$N_+=M_-=1$, see (\ref{loopsHep}). With the recently obtained two-loop heptagon \textit{function} \cite{Golden:2014xqf} we could immediately confirm our predictions for 
\beq
\mathcal{W}_{7}^{\textrm{gluons } +,-} =  \sum_{a, b \geqslant 1} e^{ia\phi_1-ib\phi_2}\int \frac{du\, dv}{(2\pi)^2} \mu_{a}(u)P_{a|{-b}}(-u|v+i0)\mu_{b}(v) e^{ip_a\sigma_{1}+ip_b\sigma_2-E_{a}\tau_1-E_{b}\tau_2} + \ldots\, , \la{Wpm}
\eeq
with all $\pi$'s included. (With the two-loop heptagon symbol \cite{simonHep}, we would be insensitive to such factors.) It would be interesting to perform this check which would probe at once all the (leading order expressions for the) helicity violating transitions $P_{a|-b}$.

More generally, from a data extraction point of view, considerably less is known about higher-loop amplitudes with $n\ge 7$ edges. Can the general~$n$ bootstrap, based on the study of cluster coordinates and associated polylogarithms \cite{Golden:2013xva} be upgraded further -- specially at higher loops -- by supplementing it with OPE boundary data, in a similar fashion to the hexagon program?
The game we played above, for instance, can be repeated for any $n$ rather straightforwardly. Using the gluonic transitions alone, it is now possible to predict the contributions $\mathcal{W}_n^{\text{gluons } a_1,\dots,a_{n-5}}$ where $a_i=\pm$. They can be derived from $\mathcal{W}_n$ by taking the limit where all the $n-5$ OPE times $\tau_i$ are large with $\tau_1 -i a_1 \phi_1,\tau_2-i a_2\phi_2,\dots$ held fixed. This simplifies $\mathcal{W}_n$ to a function of $2(n-5)$ cross-ratios (rather than $3(n-5)$) which, nevertheless, still captures a big chunk of the complete result, as illustrated above. It would be fascinating to figure out how much of the full result (if anything) is left unfixed after imposing the constraints arising from  these various double scaling limits together with the several symmetries of the problem. Of course, making this route practical would mean developing the technology for the analytical evaluation of the various OPE integrals and sums that define $\mathcal{W}_n^{\text{gluons } a_1,\dots,a_{n-5}}$. 

{Yet another option for comparing our predictions with perturbative data would be to understand whether these scaling limits can be performed already at the integrand level (or its spectral deformation~\cite{spectralstuff}) and whether it helps simplifying the resulting loop integrations. Since the integrand is very well understood to all loops~\cite{ArkaniHamed}, this could provide valuable data for the maximal helicity pieces.}
$$
\star
$$

All in all, everything seems to be working pristinely in the gluonic realm. The hope is to encounter the same good fortune when all other excitations are added back in the game but this is a  longer story.

\subsection*{Acknowledgements} 
We thank J.~Caetano, L.~Cordova, L.~Dixon, J.~Pennington, M.~von Hippel for very useful discussions and comments on the draft. This research was supported in part by Perimeter Institute for Theoretical Physics. Research at Perimeter Institute is supported by the Government of Canada through Industry Canada and by the Province of Ontario through the Ministry of Research and Innovation.  A.S was supported in part by U.S. Department of Energy grant DE- SC0009988.

\appendix

\section{Transitions and Measures} \la{weakA}

In this appendix we summarize the expressions for the transitions, measures and dispersion relations in the gluonic sector. We shall present all these results using the matrix notation introduced in \cite{data,2pt}. This way of writing is ideally suited to the numerical implementation and, most of all, to the weak coupling evaluation. 

\subsection{Summary of the Results}\label{summary}
We start by defining the matrix
\beq
\mathcal{M}\equiv\!  \(\! \begin{array}{ccccc}
+1 & 0 & 0 & 0 & \cdots \\ 
0 & -2 & 0 & 0 & \cdots \\
0 & 0 & +3 & 0 & \cdots \\ 
0 & 0 & 0 & -4 & \cdots \\
\vdots & \vdots &\vdots &\vdots & \ddots \end{array} \!\)   \cdot \[\mathbb{I}+\int\limits_0^\infty\!\! \frac{2 \,dt}{t(e^t-1)} \(\!\begin{array}{rrrrr}
J_1 J_1 & 2 J_1 J_2 & 3 J_1 J_3 & 4 J_1 J_4 & \cdots \\
 -J_2 J_1 & 2 J_2J_2 & -3 J_2 J_3 & 4 J_2 J_4& \cdots \\
 J_3 J_1 & 2 J_3 J_2 & 3 J_3J_3 & 4 J_3 J_4 & \cdots\\
 -J_4 J_1 & 2 J_4 J_2 & -3 J_4 J_3 & 4 J_4J_4 & \cdots\\
 \vdots & \vdots &\vdots &\vdots & \ddots 
\end{array}\! \)\]^{-1} \la{calM}
\eeq
where $J_i\equiv J_i(2gt)$ is the $i$-th Bessel function of the first kind and $\mathbb{I}$ is the identity matrix. In theory, $\mathcal{M}$ is an infinite (symmetric) matrix.%
\footnote{In the notations of~\cite{data} we have $\mathcal{M} = \mathbb{Q}\cdot (\mathbb{I}+\mathbb{K})^{-1}$ where $\mathbb{Q}$ stands for the first factor in~(\ref{calM}) and $\mathbb{K}$ for the second term in the square brackets. Since $\mathbb{Q}\cdot \mathbb{K} = \mathbb{K}^{t}\cdot \mathbb{Q}$ and $\mathbb{Q} = \mathbb{Q}^t$ we verify that $\mathcal{M} = \mathbb{Q}\cdot (\mathbb{I}-\mathbb{K}+\mathbb{K}^2-\ldots ) = (\mathbb{I}-\mathbb{K}^t+\mathbb{K}^{t\, 2}-\ldots )\cdot \mathbb{Q}^t  = \mathcal{M}^t$ is indeed symmetric.} In practice, it can be truncated to finite size for numerical evaluation~\cite{Benna} and the truncation will become exact in perturbation theory, thanks to the weak coupling scaling $J_i =O(g^i)$ of the Bessel functions.

This matrix is universal, in the sense that it enters into the transitions, S-matrices, measures and dispersion relations of \textit{all} the flux tube excitations. It stands for the inverse kernel of the BES equation~\cite{BES}, in the representation given in~\cite{Klebanov}, and as such appears naturally when exploring the physics around the GKP flux tube (see~\cite{BenDispPaper,withAdam,data,2pt} for various illustrations). 

The next inputs are the two (non-universal) vectors $\kappa_a(u)$ and $\tilde \kappa_a(u)$. They depend on the rapidity $u$ and type $a$ of the flux-tube excitation under consideration and their general expression was given in~\cite{BenDispPaper}. In the case of interest here $a$ is a non-zero integer labelling the gluonic excitation, as done in~(\ref{fnotation}), and, following the conventions of~\cite{data,2pt}, we get:
\beqa
\tilde{\kappa}_a = \!\int\limits_0^\infty\!\!\frac{dt}{t(e^t-1)} \sin(u t)\! \(\!\begin{array}{c} 
e^{t-|a| t/2} J_1 \\
-\, e^{-|a| t/2} J_2 \\
 e^{t-|a| t/2} J_3\\
-\, e^{-|a| t/2} J_4\\
\vdots 
\end{array}\!\!\) \, , \,\,\, {\kappa}_a = \!\int\limits_0^\infty\!\!\frac{dt}{t(e^t-1)} \(\!\begin{array}{c} 
(J_0 -e^{-|a| t/2} \cos(u t) )J_1 \\
(J_0-e^{t-|a| t/2} \cos(u t) )J_2 \\
(J_0 -e^{-|a| t/2} \cos(u t) )J_3 \\
(J_0-e^{t-|a| t/2} \cos(u t) )J_4\\
\vdots 
\end{array}\!\!\)\, . \la{kappas}
\eeqa
In those terms, the dispersion relation for the gluonic excitation $a$ can be written as
\beq
E_a(u)=|a|+4g \[\mathcal{M}\cdot \kappa_a(u)\]_1 \,, \qquad p_a(u)=2u-4g \[\mathcal{M}\cdot \tilde\kappa_a(u)\]_1\, ,
\eeq
where $\[\dots \]_1$ stands for the first element of the vector in brackets.

The important step towards constructing measures and transitions is to form the functions
\beqa
f_1^{(a,b)}(u,v) = 2 \,\tilde \kappa_{a}(u) \cdot \mathcal{M} \cdot \kappa_b(v)\, , \qquad \qquad  f_2^{(a,b)}(u,v) = 2 \, \kappa_{a}(u) \cdot \mathcal{M} \cdot \tilde\kappa_b(v) \,,\la{f1234}\\
f_3^{(a,b)}(u,v)=2 \,\tilde \kappa_{a}(u) \cdot \mathcal{M} \cdot \tilde \kappa_b(v)\, , \qquad \qquad  f_4^{(a,b)}(u,v)=2 \, \kappa_{a}(u) \cdot \mathcal{M} \cdot \kappa_b(v) \,. \nn
\eeqa
We note that, due to the symmetry of $\mathcal{M}$, they are not all independent and satisfy 
\beq
f_1^{(a,b)}(u,v) = f_2^{(b,a)}(v,u)\, , \qquad f_{3, 4}^{(a,b)}(u,v) = f_{3, 4}^{(b,a)}(v,u)\, . \la{f-sym}
\eeq
This set of functions controls the most non-trivial part of the transitions and measures, which can be written as
\beq
P_{a|b}(u|v)=F_{a,b}(u,v) \,e^{i f_2^{(a,b)}(u,v)-i f_1^{(a,b)}(u,v)+f_4^{(a,b)}(u,v)-f_3^{(a,b)}(u,v)} \,, \la{PabPhys}
\eeq
and 
\beq
\mu_{a}(u)=F_{a}(u) \,e^{f_3^{(a,b)}(u,u)-f_4^{(a,b)}(u,u)} \,.
\eeq
The only remaining ingredients are the prefactors $F_{a,b}$ and $F_a$, which are known explicitly. For $F_{a,b} \equiv F_{-a, -b}$ we should precise whether $a$ and $b$ have same or opposite signs. Taking $a,b>0$, we find
\beqa
&&\!\!\!\!\!\!\!\!\!\!\!\!F_{a,b}(u,v)=\sqrt{(x^{[+a]}y^{[-b]}-g^2)(x^{[-a]}y^{[+b]}-g^2)(x^{[+a]}y^{[+b]}-g^2)(x^{[-a]}y^{[-b]}-g^2)} \la{Fab} \\
&&  \times\, \frac{(-1)^b \Gamma(\frac{a-b}{2}+iu-iv) \Gamma(\frac{a+b}{2}-iu+iv){e^{\int_0^\infty \frac{dt (J_0(2gt)-1)}{t(e^t-1)} (J_0(2gt)+1-e^{-at/2-iu t}-e^{-bt/2+iv t})}}}{g^2 \Gamma(1+\frac{a}{2}+iu) \Gamma(1+\frac{b}{2}-iv) \Gamma(1+\frac{a-b}{2}-iu+iv)} \nn \,,
\eeqa
and
\beqa
&&F_{a,-b}(u,v)=\frac{1}{\sqrt{(1-\frac{g^2}{x^{[+a]} y^{[-b]}})(1-\frac{g^2}{x^{[-a]} y^{[+b]}})(1-\frac{g^2}{x^{[+a]} y^{[+b]}})(1-\frac{g^2}{x^{[-a]} y^{[-b]}})}} \times \\
 &&  \nn\qquad\qquad\qquad \times\,   \frac{\Gamma(1+\frac{a}{2}+iu) \Gamma(1+\frac{b}{2}-iv){e^{\int_0^\infty \frac{dt (J_0(2gt)-1)}{t(e^t-1)} (J_0(2g t)+1-e^{-at/2-iu t}-e^{-bt/2+iv t})}}}{\Gamma(1+\frac{a+b}{2}+iu-iv)} \, ,
\eeqa
with $x^{[\pm a]}=x(u\pm ia/2), y^{[\pm b]}=x(v\pm ib/2)$ and $x(u)$ as defined in~(\ref{zu}). Finally, for the measure prefactor $F_{a} = F_{-a}$, we have ($a>0$) 
\beqa
F_a(u)=\frac{(-1)^a g^2  \Gamma(1+\frac{a}{2}+iu)\Gamma(1+\frac{a}{2}-iu){e^{\int_0^\infty \frac{dt (J_0(2gt)-1)}{t(e^t-1)} (2e^{-at/2}\cos(u t)-J_0(2g t)-1)}}}{\Gamma(a)(x^{[+a]}x^{[-a]}-g^2)\sqrt{((x^{[+a]})^2-g^2)((x^{[-a]})^2-g^2)}} \,  ,
\eeqa
which immediately follows from the residue of the transition~(\ref{Fab}) at $u=v$ and $a=b$, as in~(\ref{measureDef}).

\subsection{Weak Coupling Expansion}

\def\a{n}
\def\b{m}

The expansion of all the quantities above in perturbation theory is straightforward, as explained in \cite{data} (see for example appendix E therein). The basic idea is that in perturbation theory we can Taylor expand the Bessel functions showing up in several places. Repeated use of the master integral
\beq
\int_{0}^{\infty}\frac{dt}{t}\frac{t^{k+1}}{e^{t}-1}(e^{iut}-\delta_{k, 0}) = (-1)^{k+1}(\psi_{k}(1-iu)-\delta_{k, 0}\psi(1))\,, \la{master}
\eeq
with $\psi(z) = \partial_{z}\log{\Gamma(z)}$ the Euler $\psi$-function and $\psi_{k}(z) = \partial_{z}^{k}\psi(z)$, is then enough to evaluate all resulting integrals.
A large part in this procedure can be done analytically as follows. One would first record the Taylor expansion of a Bessel function
\beq
J_{\a}(2z) = \sum_{k\geqslant 0}\frac{(-1)^k z^{\a+2k}}{\Gamma(\a+k+1)\Gamma(k+1)}\, ,
\eeq
and that of a product of two Bessel functions
\beq
J_{\a}(2z)J_{\b}(2z) = \sum_{k\geqslant 0}\frac{(-1)^{k} z^{\a+\b+2k}\Gamma(\a+\b+2k+1)}{\Gamma(\a+\b+k+1)\Gamma(\a+k+1)\Gamma(\b+k+1)\Gamma(k+1)}\, .
\eeq
Everything else follows from these two relations. For instance, using 
\beq
\int_{0}^{\infty}\frac{dt}{t}\frac{t^{k+1}}{e^{t}-1} = \Gamma(k+1)\zeta(k+1)\, ,
\eeq
which can be derived from~(\ref{master}), we get
\beq
\int_{0}^{\infty}\frac{dt}{t}\frac{J_{\a}(2gt)J_{\b}(2gt)}{e^{t}-1}=  \sum_{k\geqslant 0}g^{\a+\b+2k}\frac{(-1)^{k}\Gamma(\a+\b+2k+1)\Gamma(\a+\b+2k)\zeta(\a+\b+2k)}{\Gamma(\a+\b+k+1)\Gamma(\a+k+1)\Gamma(\b+k+1)\Gamma(k+1)}\, .
\eeq
This series can now be truncated in perturbation theory and used for evaluating the matrix $\mathcal{M}$ in (\ref{calM}). Finally, combining the above identities, another relation which is easily established is 
\beq
\begin{aligned}
&\int_0^{\infty}\frac{dt}{t}\frac{J_{\a}(2gt)(e^{iut}-J_0(2gt))}{e^{t}-1} = \\
&\sum_{k\geqslant 0}\frac{(-1)^{k}g^{\a+2k}}{\Gamma(\a+k+1)\Gamma(k+1)}\bigg[(-1)^{\a}\psi_{\a+2k-1}(1-iu)-\frac{\Gamma(\a+2k+1)\Gamma(\a+2k)}{\Gamma(\a+k+1)\Gamma(k+1)}\zeta(\a+2k)\bigg]\, ,
\end{aligned}
\eeq
here for $\a>1$. (This sum can also be used for $\a=1$ after replacing the term in square brackets by $(\psi(1)-\psi_{0}(1-iu))$ for $k=0$ and in this case only; that is, the summand should remain the same for $k\neq 0$.) This identity can  now be used to write down a series representation for the vectors $\kappa_a$ and $\tilde \kappa_a$ (which we can truncate in perturbation theory). 

In the end we obtain the vectors $\kappa_a, \tilde{\kappa}_a$ as linear combinations of polygamma functions and only the inverse operation in the definition of the matrix $\mathcal{M}$ has to be done by brute force. A short \verb"mathematica" code computing along these lines the function $f_1^{(a,b)}(u,v)$ in perturbation theory up to order $g^L$ is given below. All other quantities are either of the same complexity (the other three functions $f_2,f_3,f_4$) or simpler (everything else) and can be computed in a similar way.
The following code can be copy/pasted directly into a mathematica notebook: 
$$
\vspace{-2.3cm}
$$
\renewcommand{\arraystretch}{1.1}
\newcolumntype{L}[1]{>{\raggedright\let\newline\\\arraybackslash\hspace{0pt}}m{#1}}
\begin{center}
\begin{tabular}{ L{17cm} }
\verb"L=10;"\\ 
\verb"psi=PolyGamma; z[x_]=If[x==1,-psi[0, 1],Zeta[x]];" \\ 
\verb"collect=Collect[#,g]/.g^n_:>0/;n>L&;" \\
\verb"ClearAll[K]" \\ 
\verb"K[0]=IdentityMatrix[L-1];"\\
\verb"K[1]=Table[Sum[(i+j+2k)!(i+j+2k-1)!/((i+j+k)!(i+k)!(j+k)!k!)z[i+j+2k]2j" \verb"     g^(2k+i+j)(-1)^(j i+j+k),{k,0,L/2-i/2-j/2}],{i,L-1},{j,L-1}];" \\ 
\verb"K[n_]:=K[n]=K[n-1].K[1]//collect" \\
\verb"calM=DiagonalMatrix[#(-1)^(#+1)&/@Range[L-1]].Sum[(-1)^n K[n],{n,0,L/2}];"\\
\verb"o[i_] := Boole[OddQ[i]];"\\
\verb"kt[a_,u_]=Table[Sum[((-1)^k g^(i+2 k))/(2 I k!(i+k)!) (psi[i+2k-1,1+a/2-" \verb"          o[i]+I u]-psi[i+2k-1,1+a/2-o[i]-I u]),{k,0,L/2-i/2}],{i,1,L-1}];"\\
\verb"k[a_,u_]=Table[Sum[(-1)^(k+i)g^(i+2k)(2(-1)^i Binomial[i+2k,k](i+2k-1)!" \verb"         z[i+2k]-psi[i+2k-1,a/2+o[i]-I u]-psi[i+2k-1,a/2+o[i]+I u])" \verb"         /(2k!(i+k)!),{k,0,L/2-i/2}],{i,L-1}];"\\
\verb"f1[a_, b_][u_, v_] = 2 kt[a, u].calM.k[b, v] //collect"\end{tabular}
\end{center}

The end result for the leading and subleading expressions of the dispersion relation, measure and transitions for any bound state are given -- in \verb"Mathematica" syntax -- in the plain text file \verb"expansions.txt" attached to the arXiv submission. In this file we use \verb"P[a,b][u,v]" for $P_{a|b}(u|v)$ and \verb"Pb[a,b][u,v]" for $P_{a|-b}(u|v)$, with $a,b>0$.

\subsection{Leading Order Expressions}\label{lo-exp-app}

We conclude by presenting the leading order expressions of our results at weak coupling. These are obtained directly from the general formulae given in~\ref{summary} and are quoted here for the reader's convenience only. We have (for $a, b>0$)
\beqa
\nn E_a(u)=E_{-a}(u)\!\!\!&=&\!\!\! a+2g^2(\psi(1+\ft{a}{2}+iu)+\psi(1+\ft{a}{2}-iu)-2\psi(1)) + O(g^4)\,,\\
p_a(u)=p_{-a}(u)\!\!\!&=&\!\!\! 2u+2ig^2 (\psi(\ft{a}{2}+iu)-\psi(\ft{a}{2}-iu)) + O(g^4)\nn \,, \\
   \mu_a(u)=   \mu_{-a}(u)\!\!\!&=&\!\!\! (-1)^a g^2\frac{\Gamma(\frac{a}{2}+iu)\Gamma(\frac{a}{2}-iu)}{(\frac{a^2}{4}+
   u^2) \Gamma (a)}+O(g^4) \nn\,,\\
P_{a|b}(u|v)=P_{-a|-b}(u|v)\!\!\!&=&\!\!\! \frac{(-1)^b (\ft{a}{2}-iu) (\ft{b}{2}+iv) \Gamma(\frac{a-b}{2}+iu-iv) \Gamma(\frac{a+b}{2}-iu+iv)}{g^2 \Gamma(\frac{a}{2}+iu) \Gamma(\frac{b}{2}-iv) \Gamma(1+\frac{a-b}{2}-iu+iv)} +O(g^0) \nn\,,\\
P_{a|-b}(u|v)=P_{-a|b}(u|v)\!\!\!&=&\!\!\! \frac{\Gamma \left(1+\frac{a+b}{2}+iu-iv\right)}{\Gamma(1+\frac{a}{2}+i u) \Gamma(1+\frac{b}{2}-i v)} +O(g^2) \nn\,,\\
h_{a}(u)\!\!\!&=&\!\!\! \frac{u^2+\frac{a^2}{4}}{g^2} +O(g^0) \,,\qquad\qquad h_{-a}(u)=  \frac{g^2}{u^2+\frac{a^2}{4}} +O(g^4)\,,\la{leading}
\eeqa

\section{Goldstone Sheet and Fusion} \la{ApB}
In this appendix, we shall construct the several bound-state transitions and measures. In passing we will also review their dispersion relations. As explained in the text, our strategy is to obtain all information about the bound states by fusing together their constituent gluons. As also alluded above, it is crucial to perform this fusion in the proper kinematical domain. One convenient place is the half-mirror or \textit{Goldstone} sheet, which is depicted in green in figure~\ref{MirrorF}. Technically, we enter this sheet from the physical one by going through the first Zhukowsky cut in the upper half rapidity plane. 

There are thus two main steps in this construction: the fusion itself and the analytic continuation to or from the half-mirror sheet. Our starting point, in the first section \ref{first} of this appendix, will be the transitions for the lightest gluons evaluated in the half-mirror sheet. In this section, we shall fuse these transitions together and obtain, in this way, the transitions for the bound states in the half-mirror sheet. In the second section~\ref{second}, we will analytically continue these objects back to the physical sheet. In particular, the continuation of the transitions for the lightest gluons will lead to the expressions reported \cite{data} thus confirming the validity of our starting point. 

\subsection{Fusion in the Goldstone Sheet} \la{first}
We start by presenting, without proof, the expressions for the energy, momentum, measure and transitions for the gluonic excitations in the half-mirror sheet. Regardless of where we are, the dynamical information about an excitation can always be encoded in the form recalled in appendix \ref{summary}. What is needed is the matrix $\mathcal{M}$ in (\ref{calM}) and two vectors which were denoted as $\kappa$ and $\tilde \kappa$ in the physical sheet and which here are denoted as $k$ and $\tilde k$. (The relation between these two sets of vectors will be made clear in the next section.) The latter two vectors
turn out to be much simpler than their physical sheet counterparts (\ref{kappas}), and take the form 
\beqa
{k}_1(u)= \!\int\limits_0^\infty\!\frac{e^{i u t} dt}{2t}  \!\(\!\!\begin{array}{c} 
e^{-t/2} J_1 \\
e^{+t/2} J_2 \\
e^{-t/2} J_3 \\
e^{+t/2} J_4 \\
\vdots 
\end{array}\!\!\)\!\!=\!\frac{1}{2} \(\!\!\begin{array}{c} 
\frac{1}{1}\(\frac{ig}{x^{+}}\)^1 \\
\frac{1}{2} \(\frac{ig}{x^{-}}\)^2 \\
\frac{1}{3} \(\frac{ig}{x^{+}}\)^3 \\
\frac{1}{4} \(\frac{ig}{x^{-}}\)^4 \\
\vdots 
\end{array}\!\!\) \,, \,\,\, \tilde{k}_1(u)= \!\int\limits_0^\infty\!\frac{e^{i u t} dt}{2i t}  \!\(\!\!\begin{array}{c} 
-e^{+t/2} J_1 \\
+e^{-t/2} J_2 \\
-e^{+t/2} J_3 \\
+e^{-t/2} J_4 \\
\vdots 
\end{array}\!\!\)\!\!=\!\frac{1}{2i} \(\!\!\begin{array}{c} 
\frac{1}{1}\(\frac{g}{i x^{-}}\)^1 \\
\frac{1}{2} \(\frac{g}{i x^{+}}\)^2 \\
\frac{1}{3} \(\frac{g}{i x^{-}}\)^3 \\
\frac{1}{4} \(\frac{g}{i x^{+}}\)^4 \\
\vdots 
\end{array}\!\!\) \la{ks} \,,
\eeqa
with $x^{\pm} = x(u\pm \ft{i}{2})$. They allow us to write the dispersion relation of the twist-one gluon on the half-mirror sheet as
\beqa
E_1(\hat u)=1+4g [\mathcal{M}\cdot k_1(u)]_1 \,,\qquad p_1(\hat u)=i - 4g [\mathcal{M}\cdot {\tilde k}_1(u)]_1  \,. 
\eeqa
It is easily seen that the vectors $k_1$ and $\tilde k_1$ both vanish at $u\rightarrow \infty$, meaning that at this point we have $E_1=-ip_1=1$ (which holds at any coupling). This is the Goldstone point~\cite{AldayMaldacena}, alluded to before,
and the reason why we alternatively refer to the half-mirror sheet as the Goldstone sheet.%
\footnote{We also observe that $E_1\simeq 1$ and $p_1\simeq i$ everywhere on this sheet at weak coupling, meaning that we are covering a very small neighbourhood of the Goldtsone point. The situation is different at strong coupling where the Goldstone sheet covers the full strip between the real and mirror kinematics (i.e.~the strip $0< \textrm{Im}\,  \theta < \frac{\pi}{2}$ in the relativistic limit $E_{1}\simeq \sqrt{2}\cosh{\theta}, p_{1}\simeq \sqrt{2}\sinh{\theta}$.)}

It is straightforward to fuse the above expressions, following the procedure described around equation~(\ref{EP}), and obtain the bound-state dispersion relation. We immediately get (for $a>0$)
\beqa
E_a(\hat u)=a+4g [\mathcal{M}\cdot k_a(u)]_1 \,,\qquad p_1(\hat u)=ia - 4g [\mathcal{M}\cdot {\tilde k}_a(u)]_1  \,, \la{EPmirror}
\eeqa
in terms of the fused vectors
\beq
{k}_a(u)= \sum\limits_{k=1}^{a}k_1(u^{[+2k-a-1]}) =
 \!\int\limits_0^\infty\!\frac{dt}{2t}\, e^{iut} \,\frac{\sinh\frac{at}{2}}{\sinh\frac{t}{2}}   \!\(\!\!\begin{array}{c} 
e^{-t/2}J_1 \\
e^{+t/2} J_2 \\
e^{-t/2} J_3 \\
e^{+t/2} J_4 \\
\vdots 
\end{array}\!\!\)\!\!
=\!\frac{1}{2} \sum\limits_{k=1}^{a}\(\!\!\begin{array}{c} 
\frac{1}{1}\(\frac{ig}{x^{[+2k-a-0]}}\)^1 \\
\frac{1}{2} \(\frac{ig}{x^{[+2k-a-2]}}\)^2 \\
\frac{1}{3} \(\frac{ig}{x^{{[+2k-a-0]}}}\)^3 \\
\frac{1}{4} \(\frac{ig}{x^{{[+2k-a-2]}}}\)^4 \\
\vdots 
\end{array}\!\!\) \,,
\eeq
and similarly for the conjugate vector $\tilde k_a(u)=\sum_{k=1}^{a}\tilde{k}_1(u^{[+2k-a-1]})$. The integral representation is valid in the upper-half plane of the Goldstone sheet for $\textrm{Im}\, u > \frac{a}{2}$. 

We now present all the expressions for the measure, transitions and S-matrices of all the gluonic excitations. It is a simple exercise to check that all these quantities are related to each other through fusion as described in the main text.

For the S-matrix, we have ($a,b>0$)
\beq
S_{a,b}(\hat u,\hat v)= s_{a,b}(u,v) e^{-2i \hat f_1^{(a,b)}(u,v)+2i \hat f_2^{(a,b)}(u,v)}\, , \label{Sab-app}
\eeq
where the hatted functions $\hat f_1$, $\hat f_2$ are defined as in Eq.~(\ref{f1234}) with $k_a,\tilde k_a$ instead of $\kappa_a,\tilde \kappa_a$ and
\beq\label{stringy-fact}
s_{a,b}(u,v)=\frac{\Gamma(1+\ft{a+b}{2}-iu+iv)\Gamma(1+\ft{a-b}{2}+iu-iv)\Gamma(\ft{a+b}{2}-iu+iv)\Gamma(\ft{a-b}{2}+iu-iv)}{\Gamma(1+\ft{a+b}{2}+iu-iv)\Gamma(1+\ft{a-b}{2}-iu+iv)\Gamma(\ft{a+b}{2}+iu-iv)\Gamma(\ft{a-b}{2}-iu+iv)} \,.
\eeq
This last factor is very well known as it turns out to be identical to the scattering phase for magnon bound states (a.k.a.~Bethe strings) in compact XXX spin chains. Namely, for $a=b=1$ we have 
\beq
s_{1,1}(u,v)=\frac{u-v+i}{u-v-i} \qquad \text{and then} \qquad s_{a,b}(u,v)=\prod_{k=1}^{a} \prod_{j=1}^{b} s_{1,1}(u^{[+2k-a-1]},v^{[+2j-b-1]}) \,.
\eeq
This factor is directly responsible for the presence of bound-state poles in the S-matrix~(\ref{Sab-app}). Were it not there, there will be no bound states of gluons. It is absent from the S-matrix involving particles with opposite helicities and as a corollary gluons cannot bind together in this channel. The latter S-matrix is reflectionless and thus entirely controlled by the transmission phase. It reads ($a,b>0$)
\beq
S_{a,-b}(\hat u,\hat v)=e^{-2i \hat f_1^{(a,b)}(u,v)+2i \hat f_2^{(a,b)}(u,v)} \,,
\eeq
and is identical to~(\ref{Sab-app}) if not for the stringy prefactor~(\ref{stringy-fact}) missing. These S-matrices are unitary, $S_{a, b}(\hat{u}, \hat{v}) S_{b, a}(\hat{v}, \hat{u}) = S_{a, -b}(\hat{u}, \hat{v}) S_{b, -a}(\hat{v}, \hat{u}) = 1$, thanks to~(\ref{f-sym}).
Similarly, for the mirror S-matrix (defined as $S_{\star a,\pm b}(\hat u,\hat v) = S_{a,\mp b}(\hat u^\gamma,\hat v)$ with $\gamma$ the mirror map, see~\cite{MoreDispPaper,short,data,2pt}) one has
\beq
S_{\star a,b}(\hat u,\hat v) =\frac{e^{2 \hat f_3^{(a,b)}(u,v)-2\hat f_4^{(a,b)}(u,v)}}{h_{a,b}(u,v)}  \,, \qquad S_{\star a,-b}(\hat u,\hat v)=\frac{e^{2 \hat f_3^{(a,b)}(u,v)-2\hat f_4^{(a,b)}(u,v)}}{h_{b,a}(v,u)}  \,, 
\eeq
with 
\beq
h_{a,b}(u,v)=(-1)^b\frac{\Gamma(1+\ft{a+b}{2}-iu+iv)\Gamma(\ft{a-b}{2}+iu-iv)}{\Gamma(1+\ft{a-b}{2}-iu+iv)\Gamma(\ft{a+b}{2}+iu-iv)}\, .  \la{hab}
\eeq
We notice that, due to~(\ref{f-sym}), they satisfy the relation $S_{\star a, b}(\hat{u}, \hat{v}) = S_{\star b, -a}(\hat{v}, \hat{u})$ which is the expression of the mirror symmetry of the flux tube.

The transitions are roughly of the same complexity and read ($a,b>0$)
\beqa
P_{a|b}(\hat u|\hat v)&=&h_{a,b}(u,v) \sqrt{\frac{(1-g^2/x^{[+a]}y^{[+b]})(1-g^2/x^{[-a]}y^{[-b]})}{(1-g^2/x^{[+a]}y^{[-b]})(1-g^2/x^{[-a]}y^{[+b]})}}  \exp( \phi_{a,b}(u,v))\,, \la{PabMirror}\\
P_{-a|b}(\hat u|\hat v)&=&\sqrt{\frac{(1-g^2/x^{[+a]}y^{[-b]})(1-g^2/x^{[-a]}y^{[+b]})}{(1-g^2/x^{[+a]}y^{[+b]})(1-g^2/x^{[-a]}y^{[-b]})}}  \exp( \phi_{a,b}(u,v))\,, \la{barPabMirror}
\eeqa
with 
\beq
\phi_{a,b}(u,v)=-i \hat f_1^{(a,b)}(u,v) +i \hat f_2^{(a,b)}(u,v) -\hat f_3^{(a,b)}(u,v)+\hat f_4^{(a,b)}(u,v)  \,. \la{phi}
\eeq
One easily observes that the fundamental relation (\ref{fund}) is satisfied as a consequence of the properties~(\ref{f-sym}) of the functions $\hat{f}$ and the identity $s_{a,b}(u,v)={h_{a,b}(u,v)}/{h_{b,a}(v,u)}$
among the the stringy prefactors~(\ref{stringy-fact}) and~(\ref{hab}).
We notice that, for $a=b=1$, we have $h_{1,1} = (u-v+i)/(u-v)$. This makes manifest that $P_{1|1}$ has a simple pole at $u=v$ and a simple zero at $u=v-i$. (The remaining factor in~(\ref{PabMirror}) cannot alter this conclusion.) As explained in Section~\ref{fusion-trans}, this zero is important for the consistency of the fusion ansatz~(\ref{Pab}). Here we verify explicitly, and for an arbitrary bound state, that the fusion of $h_{1,1}$, which results in the prefactor~(\ref{hab}), does indeed lead to transitions with the expected behaviour in the square limit. Namely, the prefactor $h_{a, b}$ has a simple pole at $u=v$ (for $a=b$) and so does $P_{a|b}$. The residue at this pole gives the square measure ($a>0$)
\beq
\mu_a(\hat u)= \frac{(-1)^a (1-g^2/x^{[+a]}x^{[-a]})}{a\sqrt{(1-g^2/(x^{[+a]})^2)(1-g^2/(x^{[-a]})^2)}} \exp(-\phi_{a,a}(u,u))\,,
\eeq
which concludes the summary of the gluonic quantities in the half-mirror sheet. 

\subsection{Back to the Physical Sheet} \la{second}
In the previous section we presented the expressions for the various dynamical quantities in the Goldstone sheet, where the fusion pattern for the bound states was at each step manifest. In this section we shall explain how to analytically continue these quantities back to the physical sheet and derive the results summarized in appendix~\ref{summary}. This analysis will check the premises of the fusion procedure on the half-mirror sheet (i.e.~the correctness of the twist-one expressions we started with in appendix~\ref{first}) and simultaneously establish that the results on the physical sheet reported in appendix~\ref{summary} encode properly (though not manifestly) the fusion relations among the bound states. 

The main identity which underlies the analytic continuation through basically any cut involves introducing a new vector $\delta(u)$ and is given by\footnote{Our discussion here is closely related to the analytical continuation of the fermion transitions from the large to small physical sheet as described in Appendix C of \cite{2pt}.}
\beqa
\la{delta}\!\!\!&&\!\!\!\!\!\!\!\!\!\!\!\!\delta(u) \equiv \[\mathbb{I}+\int\limits_0^\infty\!\! \frac{2 \,dt}{t(e^t-1)} \(\!\begin{array}{rrrrr}
J_1 J_1 & 2 J_1 J_2 & 3 J_1 J_3 & 4 J_1 J_4 & \cdots \\
 -J_2 J_1 & 2 J_2J_2 & -3 J_2 J_3 & 4 J_2 J_4& \cdots \\
 J_3 J_1 & 2 J_3 J_2 & 3 J_3J_3 & 4 J_3 J_4 & \cdots\\
 -J_4 J_1 & 2 J_4 J_2 & -3 J_4 J_3 & 4 J_4J_4 & \cdots\\
 \vdots & \vdots &\vdots &\vdots & \ddots 
\end{array}\! \)\]\!\cdot\! \frac{1}{2}\! \(\!\!\begin{array}{c} 
\color{red}\frac{i^2}{1}((\frac{g}{x})^1+(\frac{x}{g})^1) \\
\color{blue}\frac{i^2}{2}((\frac{g}{x})^2+(\frac{x}{g})^2) \\
\color{red}\frac{i^4}{3}((\frac{g}{x})^3+(\frac{x}{g})^3) \\
\color{blue}\frac{i^4}{4}((\frac{g}{x})^4+(\frac{x}{g})^4) \\
\vdots 
\end{array}\!\!\)\!\! \la{identityPsi}\\&&\!\!\!\!\!\!
=\frac{1}{2} \(\!\!\begin{array}{c} 
\color{red}\frac{i^2}{1}((\frac{g}{x})^1+(\frac{x}{g})^1) \\
\color{blue}\frac{i^2}{2}((\frac{g}{x})^2+(\frac{x}{g})^2) \\
\color{red}\frac{i^4}{3}((\frac{g}{x})^3+(\frac{x}{g})^3) \\
\color{blue}\frac{i^4}{4}((\frac{g}{x})^4+(\frac{x}{g})^4) \\
\vdots 
\end{array}\!\!\)-\! \int\limits_0^\infty \frac{dt}{t(e^t-1)} \(\!\!\begin{array}{c} 
\color{blue}J_1(J_0-\cos(ut)) \\
\color{blue}J_2(J_0-\cos(ut))\\
\color{blue}J_3(J_0-\cos(ut))\\
\color{blue}J_4(J_0-\cos(ut))\\
\vdots 
\end{array}\!\!\)-\! \int\limits_0^\infty \frac{dt}{t(e^t-1)}   \(\!\!\begin{array}{c} 
\color{red}+J_1 \sin(u t) \\
\color{red}-J_2 \sin(u t) \\
\color{red}+J_3 \sin(u t) \\
\color{red}-J_4 \sin(u t) \\
\vdots 
\end{array}\!\!\) \!. \nn
\eeqa
We can further decompose this identity into its even or odd part with respect to $u \to -u$. For the even part $\delta_+$, for instance, we would drop all the odd components marked in red in this equation and keep the blue part only. We note that, by construction, the action of $\mathcal{M}$, as defined in (\ref{calM}), on the vector $\delta$ (or on its odd and even parts, $\delta_-$ and $\delta_+$) is trivial to evaluate. This is going to play an important role in the following.

Armed with this identity, we begin with the analytic continuation of the energy. We want to establish that ($a>0$)
\beqa\la{2-energy}
E_a(\hat u)=a+4g [\mathcal{M}\cdot k_a(u)]_1 \qquad \text{and} \qquad E_a(u)=a+4g \[\mathcal{M}\cdot \kappa_a(u)\]_1
\eeqa
are related by analytic continuation from the half-mirror sheet to the physical sheet. We start, in the half-mirror sheet, with the quantity 
\beq
\mathcal{M}\cdot k_a(u) = \mathcal{M}\cdot \[\frac{1}{2} \sum\limits_{k=1}^{a}\(\!\!\begin{array}{c} 
\frac{1}{1}\(\frac{ig}{x^{[+2k-a-0]}}\)^1 \\
\frac{1}{2} \(\frac{ig}{x^{[+2k-a-2]}}\)^2 \\
\frac{1}{3} \(\frac{ig}{x^{{[+2k-a-0]}}}\)^3 \\
\frac{1}{4} \(\frac{ig}{x^{{[+2k-a-2]}}}\)^4 \\
\vdots 
\end{array}\!\!\)=\frac{1}{2}\(\!\!\begin{array}{c} 
\frac{1}{1}\(\frac{ig}{x^{[+2-a]}}\)^1 \\
\color{blue}\frac{1}{2} \(\frac{ig}{x^{[-a]}}\)^2  \\
\frac{1}{3} \(\frac{ig}{x^{{[2-a]}}}\)^3 \\
\color{blue}\frac{1}{4} \(\frac{ig}{x^{{[-a]}}}\)^4 \\
\vdots 
\end{array}\!\!\)+\frac{1}{2} \sum\limits_{k=2}^{a}\(\!\!\begin{array}{c} 
\frac{1}{1}\(\frac{ig}{x^{[+2k-a-0]}}\)^1 \\
\frac{1}{2} \(\frac{ig}{x^{[+2k-a-2]}}\)^2 \\
\frac{1}{3} \(\frac{ig}{x^{{[+2k-a-0]}}}\)^3 \\
\frac{1}{4} \(\frac{ig}{x^{{[+2k-a-2]}}}\)^4 \\
\vdots 
\end{array}\!\!\)\] \la{step1}
\eeq
and recall that we are dealing here with infinite matrix and vector. The last point is of no concern as long as the components of the vector $k_a(u)$ remains sufficiently bounded or, even better, if the components of this vector become smaller and smaller as the vector index increases. The latter property is easily seen to hold true in~(\ref{step1}) provided we stay away from the cuts, which ensures that $|x|>g$. This is, however, no longer the case if we cross one of these cuts, which implies that the corresponding Zhukowsky variable flips (inside the `unit' disk) such that $|x|<g$ at the end. This clearly generates an uncomfortable growth of the vector components and makes the analytic continuation of~(\ref{step1}) potentially dangerous. The proper way of handling this situation involves performing the resummation of the large part of the vector as we now explain.

In the present case, the goal is to enter the cut of $x^{[-a]}$. 
Doing so, the terms in blue in (\ref{step1}) get flipped, i.e.~$\(g/x^{[-a]}\)^n\to \(x^{[-a]}/g\)^n$, and thence display, after crossing the cut, exponential growth with $n$ and large behaviour at small $g$. To proceed further we shall fully resum their contributions. Here is where the identity for $\delta$ (or rather for $\delta_+$) comes in.  
Precisely, we can sit right on top of the Zhukoswky cut and add and subtract $\delta_+(u^{[-a]})$ to $k_a(u)$ to get
\beq
\mathcal{M}\cdot k_a(u)=\mathcal{M}\cdot \[k_a(u)-\delta_+(u^{[-a]})\]+\mathcal{M}\cdot\delta_+(u^{[-a]}) \,. \la{add-sub}
\eeq
Several nice things happen here. First, the combination in square bracket is now free of any large contributions and can thus be directly continued down across the cut anywhere into the physical sheet. Second, the last term, which now contains all the bad terms, can be trivially evaluated using the definition of $\delta_+$. Importantly, it yields a vector with even components only -- see terms in blue in the first line in (\ref{delta}) -- and therefore it does not contribute to the energy~(\ref{2-energy}). As such, to establish the desired relation between the energies in the half-mirror and physical sheet we are left with the task of showing that the square bracket in~(\ref{add-sub}) is equal to~$\kappa_a(u)$.
Combining the expressions for $k_a$ and $\delta_+$, we find that the square bracket in question is equal to
\beq
 \!\int\limits_0^\infty\!\frac{dt}{2t}\, e^{it u} \,\frac{\sinh\frac{at}{2}}{\sinh\frac{t}{2}}   \!\(\!\!\begin{array}{c} 
e^{-t/2}J_1 \\
e^{+t/2} J_2 \\
e^{-t/2} J_3 \\
e^{+t/2} J_4 \\
\vdots 
\end{array}\!\!\)
-\frac{1}{2} \(\!\!\begin{array}{c} 
0\\
\color{blue}\frac{i^2}{2}((\frac{g}{x^{[-a]}})^2+(\frac{x^{[-a]}}{g})^2) \\
0 \\
\color{blue}\frac{i^4}{4}((\frac{g}{x^{[-a]}})^4+(\frac{x^{[-a]}}{g})^4) \\
\vdots 
\end{array}\!\!\)+\! \int\limits_0^\infty \frac{dt}{t(e^t-1)} \(\!\!\begin{array}{c} 
\color{blue}J_1(J_0-\cos(u^{[-a]}t)) \\
\color{blue}J_2(J_0-\cos(u^{[-a]}t))\\
\color{blue}J_3(J_0-\cos(u^{[-a]}t))\\
\color{blue}J_4(J_0-\cos(u^{[-a]}t))\\
\vdots 
\end{array}\!\!\)\, . \la{voila}
\eeq
On the cut, the middle vector also admits an integral representation using 
\beq 
\frac{(-1)^{n}}{4n}\Big(\Big(\frac{g}{x^{[-a]}}\Big)^{2n}+\Big(\frac{x^{[-a]}}{g}\Big)^{2n}\Big)=\int\limits_0^\infty \frac{dt}{t} J_{2n} \cos(u^{[-a]}t) \, . \eeq
Adding these three contributions up under the integral sign, we get the vector
\beq
\int\limits_0^\infty\!\!\frac{dt}{t(e^t-1)} \(\!\begin{array}{c} 
(J_0 -e^{-a t/2} \cos(u t) )J_1 \\
(J_0-e^{t-a t/2} \cos(u t) )J_2 \\
(J_0 -e^{-a t/2} \cos(u t) )J_3 \\
(J_0-e^{t-a t/2} \cos(u t) )J_4\\
\vdots 
\end{array}\!\!\)  \la{finalKappa}
\eeq
which coincides precisely with $\kappa_a(u)$ as given in~(\ref{kappas}). Importantly, we note that in the form~(\ref{finalKappa}) we can now safely continue down anywhere in the physical sheet. In particular, in contrast with the half-mirror sheet, in the physical sheet there are no cuts between $\text{Im}(u)=a/2$ and $\text{Im}(u)=-a/2$ and the representation (\ref{finalKappa}) is valid anywhere within this strip. 

The analysis for the momentum reproduces almost verbatim the one for the energy. The only significant difference is that in case the additional term $\mathcal{M}\cdot \delta_-(u^{[-a]})$ has odd components. For evaluating the momentum we are only interested in the first component which reads
\beq \[\mathcal{M}\cdot \delta_-(u^{[-a]})\]_1=-\frac{1}{2}(g/x^{[-a]}+x^{[-a]}/g)=-u^{[-a]}/(2g)\, .
\eeq
This addition accommodates precisely for the difference between the first terms in the expressions 
\beq
p_a(\hat u)=ia-4g [\mathcal{M}\cdot \tilde k_a(u)]_1 \qquad \text{and} \qquad p_a(u)=2u-4g \[\mathcal{M}\cdot \tilde\kappa_a(u)\]_1\,,
\eeq
for the momentum in either sheet. 

It is not any harder to carry out the analysis for the transitions. Suppose we start with a transition $P_{a|b}(\hat u,\hat v)$ involving two gluonic excitations of the same helicity with rapidities on the half-mirror sheet, as given in~(\ref{PabMirror}), and analytically continue each of its arguments to the physical sheet to obtain $P_{a|b}(u, v)$. The only factor which poses any challenge is the exponent $\phi_{a,b}(u,v)$ in (\ref{phi}), since the remaining prefactors in (\ref{PabMirror}) are explicit functions of the rapidities that can be analytically continued straight away. The latter exponent is composed of four functions which can all be continued along the same lines. For the sake of clarity, we shall focus here on one of them, say
\beq
\hat f_1^{(a, b)}(u,v)=2\, \tilde k_a(u) \cdot \mathcal{M} \cdot k_b(v)\, ,
\eeq
and proceed as before by moving both rapidities (from above) towards the upper Zhukowsky cuts (which is $x^{[-a]}$ for $u$ and $y^{[-b]}$ for $v$). When sitting on the cuts, we can add and subtract a few $\delta$'s, as done in~(\ref{add-sub}), to write 
\beqa\la{f-hatf}
\hat f_1^{(a,b)}(u,v)&=&2\, \tilde \kappa_a(u) \cdot \mathcal{M} \cdot \kappa_b(v) + \\&+& 2 \, \kappa_b(v)\cdot \mathcal{M} \cdot  \delta_-(u^{[-a]})+ 2 \,\kappa_a(u) \cdot \mathcal{M} \cdot \delta_+(v^{[-b]})+2 \, \delta_-(u^{[-a]})\cdot \mathcal{M} \cdot  \delta_+(v^{[-b]})\, ,\nn
\eeqa
where, using that $\mathcal{M}$ is symmetric, we flipped the right and left vectors in the first term. We now immediately notice that the first line is precisely the un-hatted function $f_1^{(a,b)}(u,v)$ in the physical sheet~(\ref{f1234}). It means that $f_1$ is the analytical continuation of $\hat  f_1$ up to the few extra terms also present in~(\ref{f-hatf}). This was to be expected since, after all, the prefactors in (\ref{PabMirror}) and (\ref{PabPhys}) are quite different looking and this difference comes precisely from these minor additional terms. What is important is that, as emphasized earlier, the action of $\mathcal{M}$ on $\delta_\pm$ can be evaluated explicitly and thence these extra terms can be turned into definite expressions of the rapidities. Indeed, it is now a straightforward (although a bit tedious) exercise to show that all the additional terms generated from $\hat f_1,\hat f_2,\hat f_3,\hat f_4$ (together with the analytic continuation of the prefactor in (\ref{PabMirror})) nicely combine to yield the prefactor in the physical sheet appearing in (\ref{PabPhys}).\footnote{The sort of manipulations involved here are similar to the ones performed in appendix C of \cite{2pt} to which the reader is referred.} The check of this statement for the lightest excitations is particularly key since it justifies the starting point of the analysis performed in appendix~\ref{first}. 

\section{Two-loop NMHV Hexagon in the All-Minus Sector} \la{W1111Ap}
Here we quote the symbol for the two-loop super Wilson loop component $\mathcal{W}^{(1111)}$ in the scaling limit where $\tau \to \infty$ with $\tau+ i \phi$ held fixed as extracted from the literature. This double scaling limit isolates the contribution of negative helicity gluons which, in the present case, first show up at two loops. (The only important subtlety which we need to keep in mind is the conversion rule between the ratio function $\mathcal{R}^{(1111)}$ -- which is normally the object of study in the literature -- and our super-loop finite ratio  $\mathcal{W}^{(1111)}$, see (\ref{WR}).) To save space we introduce the convenient notations $y=e^\sigma$, $x=e^{-\tau-i \phi}$ and combinations thereof $z=1+x y+ y^2$, $w=1+x y$ and $u=x+y$. In this alphabet the symbol takes the simple form \\ \\ 
\verb"sym=2s[y,w,w,x]+2s[y,w,w,y]+4s[y,w,x,x]-4s[y,w,y,y]-4s[y,y,w,x]-4s[y,y,w,y]-"\\
\verb"s[z,u,u,x]+s[z,u,u,y]-2s[z,u,x,x]+s[z,u,y,x]+s[z,u,y,y]-s[z,w,w,x]-s[z,w,w,y]-"
\verb"2s[z,w,x,x]+2s[z,w,y,y]-s[z,y,u,x]+s[z,y,u,y]+2s[z,y,w,x]+2s[z,y,w,y]+"\\
\verb"2s[z,y,x,x]+s[z,y,y,x]-3s[z,y,y,y]" \\ \\ 
where we use \verb"s[a,b,c,d]" to denote the symbol $a\otimes b \otimes c \otimes d$. (This expression can be directly copied into \verb"mathematica".) 
We can now convert this symbol into a series expansion around $y,x=0$ for the corresponding \textit{function} using the algorithm developed in \cite{Dixon:2014voa} (which captures all the information up to zeta values). When doing so we observe a perfect agreement with our prediction~(\ref{WNMHVexpansion}). One further check we did is to test the relation (\ref{reduction}) at symbol level.

\end{document}

%% file: Fundamental.pdf_tex
\begingroup%
  \makeatletter%
  \providecommand\color[2][]{%
    \errmessage{(Inkscape) Color is used for the text in Inkscape, but the package 'color.sty' is not loaded}%
    \renewcommand\color[2][]{}%
  }%
  \providecommand\transparent[1]{%
    \errmessage{(Inkscape) Transparency is used (non-zero) for the text in Inkscape, but the package 'transparent.sty' is not loaded}%
    \renewcommand\transparent[1]{}%
  }%
  \providecommand\rotatebox[2]{#2}%
  \ifx\svgwidth\undefined%
    \setlength{\unitlength}{456.31723633bp}%
    \ifx\svgscale\undefined%
      \relax%
    \else%
      \setlength{\unitlength}{\unitlength * \real{\svgscale}}%
    \fi%
  \else%
    \setlength{\unitlength}{\svgwidth}%
  \fi%
  \global\let\svgwidth\undefined%
  \global\let\svgscale\undefined%
  \makeatother%
  \begin{picture}(1,0.48108985)%
    \put(0,0){\includegraphics[width=\unitlength]{Fundamental.pdf}}%
    \put(0.45687053,0.25251369){\color[rgb]{0,0,0}\makebox(0,0)[lb]{\smash{${\color{black!60}\phi}$}}}%
    \put(0.52578252,0.31913401){\color[rgb]{0,0,0}\makebox(0,0)[lb]{\smash{${\color{black!60}\psi}$}}}%
    \put(0.59590917,0.38926067){\color[rgb]{0,0,0}\makebox(0,0)[lb]{\smash{$F$}}}%
    \put(0.72564348,0.38926067){\color[rgb]{0,0,0}\makebox(0,0)[lb]{\smash{$DF$}}}%
    \put(0.86239045,0.38926067){\color[rgb]{0,0,0}\makebox(0,0)[lb]{\smash{$D^2\!F$}}}%
    \put(0.38552922,0.17888071){\color[rgb]{0,0,0}\makebox(0,0)[lb]{\smash{${\color{black!60}\bar\psi}$}}}%
    \put(0.31540256,0.10875406){\color[rgb]{0,0,0}\makebox(0,0)[lb]{\smash{$\bar F$}}}%
    \put(0.16463032,0.10875406){\color[rgb]{0,0,0}\makebox(0,0)[lb]{\smash{$\bar D\bar F$}}}%
    \put(0.66844466,0.469512){\color[rgb]{0,0,0}\makebox(0,0)[lb]{\smash{$\Blue{1}$}}}%
    \put(0.80869791,0.469512){\color[rgb]{0,0,0}\makebox(0,0)[lb]{\smash{$\Blue{2}$}}}%
    \put(0.94895122,0.469512){\color[rgb]{0,0,0}\makebox(0,0)[lb]{\smash{$\Blue{3}$}}}%
    \put(0.52819136,0.469512){\color[rgb]{0,0,0}\makebox(0,0)[lb]{\smash{$\Blue{2}$}}}%
    \put(0.45806471,0.00316977){\color[rgb]{0,0,0}\makebox(0,0)[lb]{\smash{$\Red{0}$}}}%
    \put(0.52819136,0.00316977){\color[rgb]{0,0,0}\makebox(0,0)[lb]{\smash{$\Red{1\over2}$}}}%
    \put(0.59831801,0.00316977){\color[rgb]{0,0,0}\makebox(0,0)[lb]{\smash{$\Red{1}$}}}%
    \put(0.73857126,0.00316977){\color[rgb]{0,0,0}\makebox(0,0)[lb]{\smash{$\Red{2}$}}}%
    \put(0.87882456,0.00316977){\color[rgb]{0,0,0}\makebox(0,0)[lb]{\smash{$\Red{3}$}}}%
    \put(0.15652013,0.00316977){\color[rgb]{0,0,0}\makebox(0,0)[lb]{\smash{$\Red{-2}$}}}%
    \put(0.29677341,0.00316977){\color[rgb]{0,0,0}\makebox(0,0)[lb]{\smash{$\Red{-1}$}}}%
    \put(0.36690006,0.00316977){\color[rgb]{0,0,0}\makebox(0,0)[lb]{\smash{$\Red{-{1\over2}}$}}}%
    \put(-0.00126493,0.00316977){\color[rgb]{0,0,0}\makebox(0,0)[lb]{\smash{$\Red{U(1)\text{ charge}}$}}}%
    \put(0.31430512,0.469512){\color[rgb]{0,0,0}\makebox(0,0)[lb]{\smash{$\Blue{\text{twist (energy)}}$}}}%
    \put(0.03840229,0.24900736){\color[rgb]{0,0,0}\makebox(0,0)[lb]{\smash{$\bf 6$}}}%
    \put(0.03840229,0.31913401){\color[rgb]{0,0,0}\makebox(0,0)[lb]{\smash{$\bf 4$}}}%
    \put(0.03840229,0.38926067){\color[rgb]{0,0,0}\makebox(0,0)[lb]{\smash{$\bf 1$}}}%
    \put(0.03840229,0.17888071){\color[rgb]{0,0,0}\makebox(0,0)[lb]{\smash{${\bf \bar 4}$}}}%
    \put(0.03840229,0.10875406){\color[rgb]{0,0,0}\makebox(0,0)[lb]{\smash{$\bf 1$}}}%
    \put(0.01415733,0.20381946){\color[rgb]{0,0,0}\rotatebox{90}{\makebox(0,0)[lb]{\smash{$SU(4) \text{ rep}$}}}}%
  \end{picture}%
\endgroup%

%% file: PFF.pdf_tex
\begingroup%
  \makeatletter%
  \providecommand\color[2][]{%
    \errmessage{(Inkscape) Color is used for the text in Inkscape, but the package 'color.sty' is not loaded}%
    \renewcommand\color[2][]{}%
  }%
  \providecommand\transparent[1]{%
    \errmessage{(Inkscape) Transparency is used (non-zero) for the text in Inkscape, but the package 'transparent.sty' is not loaded}%
    \renewcommand\transparent[1]{}%
  }%
  \providecommand\rotatebox[2]{#2}%
  \ifx\svgwidth\undefined%
    \setlength{\unitlength}{1016.10820313bp}%
    \ifx\svgscale\undefined%
      \relax%
    \else%
      \setlength{\unitlength}{\unitlength * \real{\svgscale}}%
    \fi%
  \else%
    \setlength{\unitlength}{\svgwidth}%
  \fi%
  \global\let\svgwidth\undefined%
  \global\let\svgscale\undefined%
  \makeatother%
  \begin{picture}(1,0.1854929)%
    \put(0,0){\includegraphics[width=\unitlength]{PFF.pdf}}%
    \put(-0.01672666,0.09204868){\color[rgb]{0,0,0}\makebox(0,0)[lb]{\smash{$\displaystyle P_{1|1}(u|v)\ =$}}}%
    \put(0.20512279,0.04934827){\color[rgb]{0,0,0}\makebox(0,0)[lb]{\smash{$u$}}}%
    \put(0.18381896,0.13442358){\color[rgb]{0,0,0}\makebox(0,0)[lb]{\smash{$v$}}}%
    \put(0.20555091,0.00173571){\color[rgb]{0,0,0}\makebox(0,0)[lb]{\smash{$F$}}}%
    \put(0.18193137,0.17652024){\color[rgb]{0,0,0}\makebox(0,0)[lb]{\smash{$\bar F$}}}%
    \put(0.59738119,0.09204868){\color[rgb]{0,0,0}\makebox(0,0)[lb]{\smash{$\displaystyle P_{1|-1}(u|v)\ =$}}}%
    \put(0.834977,0.04934827){\color[rgb]{0,0,0}\makebox(0,0)[lb]{\smash{$u$}}}%
    \put(0.8136731,0.13442358){\color[rgb]{0,0,0}\makebox(0,0)[lb]{\smash{$v$}}}%
    \put(0.83540506,0.00173571){\color[rgb]{0,0,0}\makebox(0,0)[lb]{\smash{$F$}}}%
    \put(0.81178548,0.17652024){\color[rgb]{0,0,0}\makebox(0,0)[lb]{\smash{$F$}}}%
    \put(0.32969314,0.09204868){\color[rgb]{0,0,0}\makebox(0,0)[lb]{\smash{$,$}}}%
    \put(0.95954735,0.09204868){\color[rgb]{0,0,0}\makebox(0,0)[lb]{\smash{$,$}}}%
  \end{picture}%
\endgroup%

%% file: mirror.pdf_tex
\begingroup%
  \makeatletter%
  \providecommand\color[2][]{%
    \errmessage{(Inkscape) Color is used for the text in Inkscape, but the package 'color.sty' is not loaded}%
    \renewcommand\color[2][]{}%
  }%
  \providecommand\transparent[1]{%
    \errmessage{(Inkscape) Transparency is used (non-zero) for the text in Inkscape, but the package 'transparent.sty' is not loaded}%
    \renewcommand\transparent[1]{}%
  }%
  \providecommand\rotatebox[2]{#2}%
  \ifx\svgwidth\undefined%
    \setlength{\unitlength}{1245.49863281bp}%
    \ifx\svgscale\undefined%
      \relax%
    \else%
      \setlength{\unitlength}{\unitlength * \real{\svgscale}}%
    \fi%
  \else%
    \setlength{\unitlength}{\svgwidth}%
  \fi%
  \global\let\svgwidth\undefined%
  \global\let\svgscale\undefined%
  \makeatother%
  \begin{picture}(1,0.22143395)%
    \put(0,0){\includegraphics[width=\unitlength]{mirror.pdf}}%
    \put(0.0370313,0.08463562){\color[rgb]{0,0,0}\makebox(0,0)[lb]{\smash{$u^{-\gamma}$}}}%
    \put(0.35686528,0.08765168){\color[rgb]{0,0,0}\makebox(0,0)[lb]{\smash{$u$}}}%
    \put(0.19797337,0.12098782){\color[rgb]{0,0,0}\makebox(0,0)[lb]{\smash{$\displaystyle =$}}}%
    \put(0.0793365,0.17988229){\color[rgb]{0,0,0}\makebox(0,0)[lb]{\smash{$v$}}}%
    \put(0.32044091,0.17988229){\color[rgb]{0,0,0}\makebox(0,0)[lb]{\smash{$v$}}}%
    \put(0.01018651,0.05583174){\color[rgb]{0,0,0}\makebox(0,0)[lb]{\smash{$\Blue{F}$}}}%
    \put(0.37443637,0.04952544){\color[rgb]{0,0,0}\makebox(0,0)[lb]{\smash{$\Red{\bar F}$}}}%
    \put(0.07955632,0.21512537){\color[rgb]{0,0,0}\makebox(0,0)[lb]{\smash{$\Blue{F}$}}}%
    \put(0.32106601,0.21512537){\color[rgb]{0,0,0}\makebox(0,0)[lb]{\smash{$\Blue{F}$}}}%
    \put(0.56371438,0.06528543){\color[rgb]{0,0,0}\makebox(0,0)[lb]{\smash{$u^{5\gamma}$}}}%
    \put(0.62500588,0.06501873){\color[rgb]{0,0,0}\makebox(0,0)[lb]{\smash{$v$}}}%
    \put(0.19477682,0.00135416){\color[rgb]{0,0,0}\makebox(0,0)[lb]{\smash{$({\bf a})$}}}%
    \put(0.61927438,0.0229889){\color[rgb]{0,0,0}\makebox(0,0)[lb]{\smash{$\Blue{F}$}}}%
    \put(0.59229724,0.0229889){\color[rgb]{0,0,0}\makebox(0,0)[lb]{\smash{$\Blue{F}$}}}%
    \put(0.76449342,0.12098782){\color[rgb]{0,0,0}\makebox(0,0)[lb]{\smash{$\displaystyle =$}}}%
    \put(0.75487378,0.00135416){\color[rgb]{0,0,0}\makebox(0,0)[lb]{\smash{$({\bf b})$}}}%
    \put(0.90574599,0.0229889){\color[rgb]{0,0,0}\makebox(0,0)[lb]{\smash{$\Red{\bar F}$}}}%
    \put(0.87876885,0.0229889){\color[rgb]{0,0,0}\makebox(0,0)[lb]{\smash{$\Blue{F}$}}}%
    \put(0.88234352,0.06473905){\color[rgb]{0,0,0}\makebox(0,0)[lb]{\smash{$v$}}}%
    \put(0.91188992,0.06473905){\color[rgb]{0,0,0}\makebox(0,0)[lb]{\smash{$u$}}}%
  \end{picture}%
\endgroup%

%% file: axiom1.pdf_tex
\begingroup%
  \makeatletter%
  \providecommand\color[2][]{%
    \errmessage{(Inkscape) Color is used for the text in Inkscape, but the package 'color.sty' is not loaded}%
    \renewcommand\color[2][]{}%
  }%
  \providecommand\transparent[1]{%
    \errmessage{(Inkscape) Transparency is used (non-zero) for the text in Inkscape, but the package 'transparent.sty' is not loaded}%
    \renewcommand\transparent[1]{}%
  }%
  \providecommand\rotatebox[2]{#2}%
  \ifx\svgwidth\undefined%
    \setlength{\unitlength}{1036.26181641bp}%
    \ifx\svgscale\undefined%
      \relax%
    \else%
      \setlength{\unitlength}{\unitlength * \real{\svgscale}}%
    \fi%
  \else%
    \setlength{\unitlength}{\svgwidth}%
  \fi%
  \global\let\svgwidth\undefined%
  \global\let\svgscale\undefined%
  \makeatother%
  \begin{picture}(1,0.24838145)%
    \put(0,0){\includegraphics[width=\unitlength]{axiom1.pdf}}%
    \put(-0.00036071,0.23959508){\color[rgb]{0,0,0}\makebox(0,0)[lb]{\smash{$\displaystyle P_{\dots,a_{i},a_{i+1},\dots}(\dots, u_{i},u_{i+1},\dots|0)=S_{a_i,a_{i+1}}(u_i,u_{i+1})\times P_{\dots,a_{i+1},a_{i},\dots}(\dots, u_{i+1},u_{i},\dots|0)$}}}%
    \put(0.04781678,0.11577218){\color[rgb]{0,0,0}\makebox(0,0)[lb]{\smash{$u_i$}}}%
    \put(0.09568114,0.11577218){\color[rgb]{0,0,0}\makebox(0,0)[lb]{\smash{$u_{i+1}$}}}%
    \put(0.66696533,0.11577218){\color[rgb]{0,0,0}\makebox(0,0)[lb]{\smash{$u_{i+1}$}}}%
    \put(0.72409375,0.11577218){\color[rgb]{0,0,0}\makebox(0,0)[lb]{\smash{$u_i$}}}%
  \end{picture}%
\endgroup%

%% file: axiom2.pdf_tex
\begingroup%
  \makeatletter%
  \providecommand\color[2][]{%
    \errmessage{(Inkscape) Color is used for the text in Inkscape, but the package 'color.sty' is not loaded}%
    \renewcommand\color[2][]{}%
  }%
  \providecommand\transparent[1]{%
    \errmessage{(Inkscape) Transparency is used (non-zero) for the text in Inkscape, but the package 'transparent.sty' is not loaded}%
    \renewcommand\transparent[1]{}%
  }%
  \providecommand\rotatebox[2]{#2}%
  \ifx\svgwidth\undefined%
    \setlength{\unitlength}{1056.15742187bp}%
    \ifx\svgscale\undefined%
      \relax%
    \else%
      \setlength{\unitlength}{\unitlength * \real{\svgscale}}%
    \fi%
  \else%
    \setlength{\unitlength}{\svgwidth}%
  \fi%
  \global\let\svgwidth\undefined%
  \global\let\svgscale\undefined%
  \makeatother%
  \begin{picture}(1,0.16968521)%
    \put(0,0){\includegraphics[width=\unitlength]{axiom2.pdf}}%
    \put(-0.00071315,0.16312003){\color[rgb]{0,0,0}\makebox(0,0)[lb]{\smash{$\displaystyle  i  \,\underset{u_2=u_1}{\rm residue}\,P_{a_1,a_2,a_3,\dots,a_N}(u_1^{2\gamma} , u_2,u_3,\dots,u_N|0)= \frac{\delta_{a_1,a_2}}{\mu_{a_1}(u_1)}\times P_{a_3,\dots,a_N}(u_3,\dots,u_N|0)\,.$}}}%
    \put(0.63684725,0.05501956){\color[rgb]{0,0,0}\makebox(0,0)[lb]{\smash{$\displaystyle\times$}}}%
  \end{picture}%
\endgroup%

%% file: axiom3.pdf_tex
\begingroup%
  \makeatletter%
  \providecommand\color[2][]{%
    \errmessage{(Inkscape) Color is used for the text in Inkscape, but the package 'color.sty' is not loaded}%
    \renewcommand\color[2][]{}%
  }%
  \providecommand\transparent[1]{%
    \errmessage{(Inkscape) Transparency is used (non-zero) for the text in Inkscape, but the package 'transparent.sty' is not loaded}%
    \renewcommand\transparent[1]{}%
  }%
  \providecommand\rotatebox[2]{#2}%
  \ifx\svgwidth\undefined%
    \setlength{\unitlength}{1669.02363281bp}%
    \ifx\svgscale\undefined%
      \relax%
    \else%
      \setlength{\unitlength}{\unitlength * \real{\svgscale}}%
    \fi%
  \else%
    \setlength{\unitlength}{\svgwidth}%
  \fi%
  \global\let\svgwidth\undefined%
  \global\let\svgscale\undefined%
  \makeatother%
  \begin{picture}(1,0.19347085)%
    \put(0,0){\includegraphics[width=\unitlength]{axiom3.pdf}}%
    \put(0.00107786,-0.00446299){\color[rgb]{0,0,0}\makebox(0,0)[lb]{\smash{$u_1^{5\gamma}$}}}%
    \put(0.49669696,0.00128888){\color[rgb]{0,0,0}\makebox(0,0)[lb]{\smash{$u_1$}}}%
    \put(-0.00135214,0.18102324){\color[rgb]{0,0,0}\makebox(0,0)[lb]{\smash{$\displaystyle P_{a_1,a_2,\dots,a_N}(u_1^{5\gamma} , u_2,\dots,u_N|0)=P_{a_2,\dots,a_N,-a_1}(u_2,\dots,u_N,u_1|0) \,.$}}}%
  \end{picture}%
\endgroup%

%% file: HexagonNMHV.pdf_tex
\begingroup%
  \makeatletter%
  \providecommand\color[2][]{%
    \errmessage{(Inkscape) Color is used for the text in Inkscape, but the package 'color.sty' is not loaded}%
    \renewcommand\color[2][]{}%
  }%
  \providecommand\transparent[1]{%
    \errmessage{(Inkscape) Transparency is used (non-zero) for the text in Inkscape, but the package 'transparent.sty' is not loaded}%
    \renewcommand\transparent[1]{}%
  }%
  \providecommand\rotatebox[2]{#2}%
  \ifx\svgwidth\undefined%
    \setlength{\unitlength}{1018.65800781bp}%
    \ifx\svgscale\undefined%
      \relax%
    \else%
      \setlength{\unitlength}{\unitlength * \real{\svgscale}}%
    \fi%
  \else%
    \setlength{\unitlength}{\svgwidth}%
  \fi%
  \global\let\svgwidth\undefined%
  \global\let\svgscale\undefined%
  \makeatother%
  \begin{picture}(1,0.45444618)%
    \put(0,0){\includegraphics[width=\unitlength]{HexagonNMHV.pdf}}%
    \put(0.65667137,0.16830841){\color[rgb]{0,0,0}\makebox(0,0)[lb]{\smash{$P^*$}}}%
    \put(0.87110922,0.07778763){\color[rgb]{0,0,0}\makebox(0,0)[lb]{\smash{$1/\mu$}}}%
    \put(0.73717323,0.36179351){\color[rgb]{0,0,0}\makebox(0,0)[lb]{\smash{$P$}}}%
    \put(0.45453033,0.23181242){\color[rgb]{0,0,0}\makebox(0,0)[lb]{\smash{$=\ \ \sum\limits_\psi$}}}%
    \put(0.28773178,0.04656203){\color[rgb]{0,0,0}\makebox(0,0)[lb]{\smash{$\Blue{F}$}}}%
    \put(0.33443657,0.13424844){\color[rgb]{0,0,0}\makebox(0,0)[lb]{\smash{$\,_1$}}}%
    \put(0.33386261,0.33777454){\color[rgb]{0,0,0}\makebox(0,0)[lb]{\smash{$\,_3$}}}%
    \put(0.2505493,0.35730357){\color[rgb]{0,0,0}\makebox(0,0)[lb]{\smash{$\,_4$}}}%
    \put(0.21555776,0.26306409){\color[rgb]{0,0,0}\makebox(0,0)[lb]{\smash{$\,_5$}}}%
    \put(0.24876329,0.15740175){\color[rgb]{0,0,0}\makebox(0,0)[lb]{\smash{$\,_6$}}}%
    \put(0.36642665,0.22800077){\color[rgb]{0,0,0}\makebox(0,0)[lb]{\smash{$\,_2$}}}%
    \put(0.6150861,0.347947){\color[rgb]{0,0,0}\makebox(0,0)[lb]{\smash{$_\psi$}}}%
    \put(0.78943313,0.21286731){\color[rgb]{0,0,0}\makebox(0,0)[lb]{\smash{$_\psi$}}}%
    \put(0.89781101,0.17517066){\color[rgb]{0,0,0}\makebox(0,0)[lb]{\smash{$_\psi$}}}%
    \put(-0.00097092,0.23181242){\color[rgb]{0,0,0}\makebox(0,0)[lb]{\smash{$\displaystyle \cW_\text{hex}^{(1111)}\ \ =$}}}%
    \put(0.89781101,0.00239432){\color[rgb]{0,0,0}\makebox(0,0)[lb]{\smash{$_\psi$}}}%
  \end{picture}%
\endgroup%

%% file: chargedtrans3.pdf_tex
\begingroup%
  \makeatletter%
  \providecommand\color[2][]{%
    \errmessage{(Inkscape) Color is used for the text in Inkscape, but the package 'color.sty' is not loaded}%
    \renewcommand\color[2][]{}%
  }%
  \providecommand\transparent[1]{%
    \errmessage{(Inkscape) Transparency is used (non-zero) for the text in Inkscape, but the package 'transparent.sty' is not loaded}%
    \renewcommand\transparent[1]{}%
  }%
  \providecommand\rotatebox[2]{#2}%
  \ifx\svgwidth\undefined%
    \setlength{\unitlength}{1300.20068359bp}%
    \ifx\svgscale\undefined%
      \relax%
    \else%
      \setlength{\unitlength}{\unitlength * \real{\svgscale}}%
    \fi%
  \else%
    \setlength{\unitlength}{\svgwidth}%
  \fi%
  \global\let\svgwidth\undefined%
  \global\let\svgscale\undefined%
  \makeatother%
  \begin{picture}(1,0.23253633)%
    \put(0,0){\includegraphics[width=\unitlength]{chargedtrans3.pdf}}%
    \put(0.05304779,0.22649317){\color[rgb]{0,0,0}\makebox(0,0)[lb]{\smash{$\Red{\bar F}$}}}%
    \put(0.02474446,0.11451047){\color[rgb]{0,0,0}\makebox(0,0)[lb]{\smash{$P^*_1(0|u)$}}}%
    \put(0.32008349,0.00129718){\color[rgb]{0,0,0}\makebox(0,0)[lb]{\smash{$\Red{\bar F}$}}}%
    \put(0.265938,0.11451047){\color[rgb]{0,0,0}\makebox(0,0)[lb]{\smash{$P^*_{-1}(u|0)$}}}%
    \put(0.18841151,0.11451047){\color[rgb]{0,0,0}\makebox(0,0)[lb]{\smash{$=$}}}%
    \put(0.65603163,0.22649317){\color[rgb]{0,0,0}\makebox(0,0)[lb]{\smash{$\Blue{F}$}}}%
    \put(0.62280599,0.11451047){\color[rgb]{0,0,0}\makebox(0,0)[lb]{\smash{$P^*_{-1}(0|u)$}}}%
    \put(0.92306733,0.00129718){\color[rgb]{0,0,0}\makebox(0,0)[lb]{\smash{$\Blue{F}$}}}%
    \put(0.87384416,0.11451047){\color[rgb]{0,0,0}\makebox(0,0)[lb]{\smash{$P^*_1(u|0)$}}}%
    \put(0.79139535,0.11451047){\color[rgb]{0,0,0}\makebox(0,0)[lb]{\smash{$=$}}}%
  \end{picture}%
\endgroup%

%% file: string.pdf_tex
\begingroup%
  \makeatletter%
  \providecommand\color[2][]{%
    \errmessage{(Inkscape) Color is used for the text in Inkscape, but the package 'color.sty' is not loaded}%
    \renewcommand\color[2][]{}%
  }%
  \providecommand\transparent[1]{%
    \errmessage{(Inkscape) Transparency is used (non-zero) for the text in Inkscape, but the package 'transparent.sty' is not loaded}%
    \renewcommand\transparent[1]{}%
  }%
  \providecommand\rotatebox[2]{#2}%
  \ifx\svgwidth\undefined%
    \setlength{\unitlength}{619.29248047bp}%
    \ifx\svgscale\undefined%
      \relax%
    \else%
      \setlength{\unitlength}{\unitlength * \real{\svgscale}}%
    \fi%
  \else%
    \setlength{\unitlength}{\svgwidth}%
  \fi%
  \global\let\svgwidth\undefined%
  \global\let\svgscale\undefined%
  \makeatother%
  \begin{picture}(1,0.66509634)%
    \put(0,0){\includegraphics[width=\unitlength]{string.pdf}}%
    \put(0.2591043,0.38343994){\color[rgb]{0,0,0}\makebox(0,0)[lb]{\smash{$i$}}}%
    \put(0.38075591,0.47386571){\color[rgb]{0,0,0}\makebox(0,0)[lb]{\smash{$\Red{i}$}}}%
    \put(0.94117295,0.63146491){\color[rgb]{0,0,0}\makebox(0,0)[lb]{\smash{$u$}}}%
    \put(0.48659647,0.33618737){\color[rgb]{0,0,0}\makebox(0,0)[lb]{\smash{$u$}}}%
    \put(0.48659647,0.62038265){\color[rgb]{0,0,0}\makebox(0,0)[lb]{\smash{$u^{[2]}$}}}%
    \put(0.48659647,0.01840533){\color[rgb]{0,0,0}\makebox(0,0)[lb]{\smash{$u^{[-2]}$}}}%
  \end{picture}%
\endgroup%

%% file: heptagonNMHV2.pdf_tex
\begingroup%
  \makeatletter%
  \providecommand\color[2][]{%
    \errmessage{(Inkscape) Color is used for the text in Inkscape, but the package 'color.sty' is not loaded}%
    \renewcommand\color[2][]{}%
  }%
  \providecommand\transparent[1]{%
    \errmessage{(Inkscape) Transparency is used (non-zero) for the text in Inkscape, but the package 'transparent.sty' is not loaded}%
    \renewcommand\transparent[1]{}%
  }%
  \providecommand\rotatebox[2]{#2}%
  \ifx\svgwidth\undefined%
    \setlength{\unitlength}{1225.86591797bp}%
    \ifx\svgscale\undefined%
      \relax%
    \else%
      \setlength{\unitlength}{\unitlength * \real{\svgscale}}%
    \fi%
  \else%
    \setlength{\unitlength}{\svgwidth}%
  \fi%
  \global\let\svgwidth\undefined%
  \global\let\svgscale\undefined%
  \makeatother%
  \begin{picture}(1,0.41957758)%
    \put(0,0){\includegraphics[width=\unitlength]{heptagonNMHV2.pdf}}%
    \put(-0.00051921,0.27683945){\color[rgb]{0,0,0}\makebox(0,0)[lb]{\smash{$\displaystyle\mathcal{W}_{7}^{(1111)\textrm{ gluons } +,+} =$}}}%
    \put(-0.00051921,0.04190344){\color[rgb]{0,0,0}\makebox(0,0)[lb]{\smash{$\displaystyle=  \sum_{a, b \geqslant 1} e^{ia\phi_1+ib\phi_2}\int \frac{du dv}{(2\pi)^2} {\color{blue} h_a(u)}\mu_{a}(u)P_{a|b}(-u|v+i0)\mu_{b}(v) e^{ip_a\sigma_{1}+ip_b\sigma_2
-E_{a}\tau_1-E_{b}\tau_2} + \ldots\, ,$}}}%
    \put(0.31942017,0.11970769){\color[rgb]{0,0,0}\makebox(0,0)[lb]{\smash{$_1$}}}%
    \put(0.38352776,0.18910543){\color[rgb]{0,0,0}\makebox(0,0)[lb]{\smash{$_2$}}}%
    \put(0.23512373,0.28088597){\color[rgb]{0,0,0}\makebox(0,0)[lb]{\smash{$_6$}}}%
    \put(0.39788496,0.33789821){\color[rgb]{0,0,0}\makebox(0,0)[lb]{\smash{$_3$}}}%
    \put(0.33523537,0.4122946){\color[rgb]{0,0,0}\makebox(0,0)[lb]{\smash{$_4$}}}%
    \put(0.25953378,0.3979374){\color[rgb]{0,0,0}\makebox(0,0)[lb]{\smash{$_5$}}}%
    \put(0.25470173,0.14383999){\color[rgb]{0,0,0}\makebox(0,0)[lb]{\smash{$_7$}}}%
  \end{picture}%
\endgroup%